%% file: arxiv.tex
\documentclass[manuscript,screen,nonacm]{acmart}
\AtBeginDocument{%
  }

\usepackage[normalem]{ulem}

\usepackage{shortcuts}
\usepackage{float}
\setcopyright{none}

\begin{document}

\title{Public Perceptions of Public Participation in Local Government Mediated through a GenAI Tool}

\author{Jason Yan}
\email{jasonyan@umich.edu}
\affiliation{%
  \institution{University of Michigan}
  \city{Ann Arbor}
  \state{Michigan}
  \country{USA}
}

\author{Yanna Krupnikov}
\email{yanna@umich.edu}
\affiliation{%
  \institution{University of Michigan}
  \city{Ann Arbor}
  \state{Michigan}
  \country{USA}
}

\author{Nicholas Weber}
\email{nmweber@uw.edu}
\affiliation{%
  \institution{University of Washington}
  \city{Seattle}
  \state{Washington}
  \country{USA}
}

\author{Sabina Tomkins}
\email{stomkins@umich.edu}
\affiliation{%
  \institution{University of Michigan}
  \city{Ann Arbor}
  \state{Michigan}
  \country{USA}
}

\renewcommand{\shortauthors}{Yan et al.}

\begin{abstract}
City council meetings offer a site for direct democracy through public comment periods. Yet, this process can lack transparency, and   research on local political participation has documented that people who
  speak at public meetings are often unrepresentative of their communities.
  However, a lack of representation does not establish whether other
  people view these participants as advocates for their concerns.
 In fact, little
  is known about how they are perceived. 
To address this gap and improve transparency into this democratic process, we built a generative AI (\genai) tool, \ourtool{}, that allows individuals to retrieve and read public comments. We deploy this tool, along with a novel survey,
in a study with 72 participants in Ann Arbor, Michigan, and Seattle, Washington,
in order to understand people's perceptions of public commenters. 
We find that participants generally viewed public commenters favorably, and that these views became more positive after using \ourtool{}. For example, the share who agreed that public commenters advocate for issues they care about rose from 49\% to 62\%. Knowledge of how public comment works in their city was moderate and also improved: the proportion of participants who correctly answered a procedural knowledge question about council meetings increased from 53\% to 74\%. Participants viewed public commenters more favorably than local government, yet their sense of being heard by local government also grew: agreement that their own voice is heard rose from 28\% to 46\%. These findings suggest that giving residents opportunities to engage with and learn from records of local political participation can foster positive civic outcomes.

\end{abstract}



\maketitle

\section{Introduction}
Public participation in local governance has the capacity to influence local policy \citep{sahn2025}. One important site for this potential influence is public comment periods -- time set aside during local government meetings for the public to speak directly to their elected officials \citep{sahn2025, einstein2019participates, yoder2020}. 
Despite their potential importance, however, many (if not most) people do not participate in public comment periods \cite{einstein2019defenders, einstein2019participates, yoder2020, collins2021meeting}. Beyond low participation rates, the considerable disparities in who participates in public comment periods -- and the consequent unrepresentativeness of the commenters -- have been well documented~\cite{einstein2019participates,martin2026participation}.  Public commenters, for example, tend to be older, disproportionately White, more likely to be male, educated and to own their homes than other residents in an area represented by the city council~\cite{einstein2019participates,martin2026participation, smith2013civic}. 

There are a number of factors underlying both the low participation and the participatory disparities.  Research has long considered the relative ``costs'' (e.g., effort, time) of participation -- suggesting that people have differential capabilities of handling costs \cite{verba1995voice}. Applying these ideas to city council meetings, researchers have suggested the possibilities of differential costs leading to differential participation \cite{einsteinetal2023}. A second factor shaping participatory behavior is people's perceptions of the meetings themselves~\cite{hibbing2002stealth,innes2004reframing} -- the possibility that people are avoiding public comment periods because they believe these are places where ``a select few show up to fight for narrow specialized group interests, even at the expense of a greater good'' \cite{collins2021meeting}. Research in this area has often focused on the role of meeting structure; showing people that meetings could include open participation and deliberation, for example, changes how people think about the local governance process~\cite{collins2021meeting,collins2022corrigendum}, while a weaker local information ecosystem has been shown to negatively impact political participation~\cite{hayes2018decline,hayes2021newshole,shaker2014dead}.

\textbf{Thus, low and non-representative participation in public commenting periods (and local politics in general) may be mitigated by providing awareness around those who participate, and transparency into the actions of local government.}
Typically, people rarely observe community opinion directly; they infer it largely from
news coverage \citep{mutz1998impersonal,mutz1997reading}, which  is also critical in holding local government accountable. However, as local news is in decline, this coverage is threatened~\cite{abernathy2020newsdeserts,hayes2021newshole}. Even where local news does persist, news coverage can amplify issues that are unrepresentative of the broader community: coverage often favors conflict and drama as being more newsworthy over routine and institutionalized proceedings that span a wide range of everyday issues %
~\cite{galtung1965structure,harcup2017news}. 

One avenue for broadening awareness and increasing transparency %
is to expose residents to the real comments spoken in these meetings with civic technology (civic tech), perhaps by leveraging recent advances in generative artificial intelligence (\genai).
This approach may be able to complement local news, while also offering unique advantages -- e.g., \genai could surface the opinions that residents are most interested in and provide a holistic view of community voices. 

A growing body of civic tech work has begun to make records of local government meetings more accessible \citep{barari2023localview,brown2021councildataproject,xu2025publicspeak,ge2025public,hu2023meetingbank,nickels2024documenters,kelly2025bridging}.

However, none of this work has investigated how exposure to these records can impact civic attitudes and knowledge. 
In particular, we theorize that providing exposure to public comments might impact
perceptions of public commenters, perceived alignment with others on local political issues, procedural knowledge about council meetings, and perceptions of local political efficacy.
The effect of seeing comments may depend on one's  civic attitudes, knowledge, and engagement; however, we are lacking relevant empirical data on these measures.
We are also lacking direction from the public about the types of information they need about their city councils.
Thus, we ask:

 \begin{description}
  \item[RQ1:] What are existing perceptions of public commenters, perceived alignment with others on  local political issues, procedural knowledge about city council meetings, and local political efficacy?
\item[RQ2:] How does using a \genai tool impact  local civic attitudes and knowledge? 
  \item[RQ3:] What information would \subs{} like provided in the future by \genai tools?
\end{description}

\paragraph{Contributions and Findings}
We make three contributions. 
The first is a novel survey instrument to collect measures of relevant civic attitudes and knowledge.
The second is \ourtool{}, a \genai-based tool we develop for surfacing and engaging with real public comments. \ourtool{} 
retrieves public comments from council meetings and summarizes them, with every statement in the summary linked to the comment that supports it.
Third, we use both the survey instrument and \ourtool{} in a study with 72 residents of Ann Arbor, Michigan, and Seattle, Washington. We find:

\begin{itemize}
    \item Most participants (72\%) agreed that public comment leads to better outcomes for residents, but only 28\% felt that their own voice is heard in local government decisions.
    \item Participants' own issue priorities closely matched the priorities they attributed to their community. %
    This contrasts with pluralistic ignorance, in which people believe their views differ from those of others~\cite{miller1987pluralistic}.
    \item After using \ourtool{}, more participants felt that their voice is heard in local government (from 28\% to 46\%) and that public commenters advocate for issues they care about (from 49\% to 62\%). %
    \item When asked what information they would want from a civic AI tool, participants most often requested information about local issues, accurate and well-sourced claims, and records of council decisions. Ten participants raised concerns about such tools, including political framing, regulation, and the loss of human interaction.
\end{itemize}

\section{Related Works}
\label{sec:related}
Our work both investigates perceptions that people hold of public commenters (and other civic attitudes and knowledge) and offers a civic tech tool for informing (and potentially correcting) those perceptions and broadening civic engagement. 
We first discuss literature on people's perceptions of those who comment during meetings. We then discuss the role of journalism in both shaping these perceptions and providing transparency into the workings of local government.
Finally, we discuss related work in civic tech.

\subsection{Perceptions of public commenters and civic engagement}  
Perceptions of the types of comments made by  fellow local residents may matter for civic engagement. People, research suggests, often misperceive where others stand. Sometimes, for example, people may believe that their own concerns
are shared by few others~\cite{miller1987pluralistic,sparkman2022false}. Another misperception may lead people to believe that extreme views are majority views, even when that is not the case \cite{bloy2026}. These types of incorrect perceptions about the views of others could be consequential. Those who believe they hold a minority view, for example, may be less willing to speak
up~\cite{noelleneumann1974spiral,glynn1997perceived,matthes2018spiral, settlecarlson2019}. 

Yet although theories of political expression suggest that both beliefs about and exposure to the voices of others should alter people's willingness to participate in political discourse \cite {settlecarlson2019, weeksetal2024}, there is relatively little consideration of these ideas in the context of public comment periods. Given the relatively low information many people have about their local governments \cite{einstein2019participates}, this is a case where exposure to the types of comments other residents make in city council meetings may prove pivotal. This type of exposure may not only clarify what is being said during public comment periods, but in some cases, even inform people that these meetings \emph{have} public comment periods in the first place. 

\subsection{Role of local journalism in informing perceptions of public commenters and city council meetings}
Residents have historically learned about city council meetings through local
journalists, although that coverage has never been exhaustive~\cite{Lacyetal1999}. Even in the case of local news,  media norms push journalists toward stories that have conflict,
negativity, and unexpectedness, so coverage tends to favor unusual or
dramatic events over routine proceedings~\cite{galtung1965structure,harcup2017news, weaver1986}. Early studies, for example, suggest that while local news did frequently cover city council meetings and decisions, journalists exercised some selection in their focus \cite{weaver1986}. 
News norms aside, however, over the past two decades, local journalism in the United States has also
contracted sharply \citep{abernathy2020newsdeserts,medill2024stateoflocalnews}. %
These
declines in local
political coverage have been linked to reduced political and civic
engagement~\cite{hayes2018decline,hayes2021newshole,shaker2014dead}. 

A direct consequence of the loss of local outlets  has been a
turn to national news -- which centers on national rather than local
politics~\cite{martin2019local,ellger2024local}.  Even if news norms guided local news coverage \cite{weaver1986}, that coverage still offered visibility into and a focus on local government. The absence of this coverage has meant a national lens applied to local politics \cite{hopkins2018}.  A reliance on national coverage -- and national political divides -- leads people to overestimate how divided the
public is~\cite{levendusky2016misperceptions}. A national perspective has also led to greater uniformity across geographically disparate local elections. Newspaper closures are
associated with more party-line voting~\cite{darr2018newspaper} and greater
support for parties further from the ideological center~\cite{ellger2024local}.
Moreover, 
local elections increasingly track national ones~\cite{moskowitz2021local}.
Residents may therefore know less about what their own community actually
thinks, and  assume either more division or more uniformity than exists.

Regardless, local journalism can serve an essential watchdog role. 
In California cities whose newspapers cut newsroom staff, mayoral elections
drew fewer challengers and became less competitive~\cite{rubado2020watchdog},
and after local newspapers close, municipal borrowing costs rise, consistent
with reduced scrutiny of local government~\cite{gao2020financing}. City
council meetings are where many local decisions are made in public, but as
coverage declines, fewer residents learn what happens there. Public comments
offer one partial window into these meetings. Public commenters often speak
about issues their local government is addressing and decisions it has made
or is considering, so what they say can indirectly reveal what their local
government is doing, as well as how residents are reacting to it.

This motivates our work. How do people perceive public commenters and their
local government? Does seeing what public commenters say at council meetings
change these perceptions, including whether people feel their own issues are
advocated for and whether they see their local government as responsive to
people like them?

\subsection{Civic Technology for Council Meetings and the Role of \genai}

Many local governments publish recordings of their council meetings online in accordance with state laws requiring openly accessible meetings (e.g., Sunshine Laws),\footnote{``Sunshine laws'' refers collectively to federal and state open-government laws. At the federal level, the Government in the Sunshine Act of 1976 covers federal agencies. Each state also has its own open meetings and public records laws. Broadly, these laws require public bodies to meet in public and make their records available. Many city councils have met these requirements by posting recordings of their meetings online (e.g., on YouTube).} and a growing body of work builds tools on these records. LocalView collects videos and transcripts of local government meetings posted to YouTube into a database for the study of local politics and policy making~\cite{barari2023localview}. The Council Data Project provides open-source software that collects, transcribes, and republishes municipal meeting data through a searchable web application~\cite{brown2021councildataproject}. The Council Data Project has also been used to assemble a dataset of more than 350 meetings of the Seattle City Council, Portland City Council, and King County Council, which was used to track how often topics such as policing, housing, and homelessness arose in council discussion over time and across councils~\cite{brown2022councilsinaction}. MeetingBank pairs city council meeting transcripts with reference summaries to support automatic meeting summarization~\cite{hu2023meetingbank}. Other work focuses on public comment specifically. PublicSpeak is a framework for identifying public comments within meeting transcripts~\cite{xu2025publicspeak}, and a separate study of public comments across Michigan characterizes the issues that members of the public raise~\cite{ge2025public}. Beyond software, the Documenters Network trains and pays community members to take notes at public meetings, producing records that are more readable than official minutes~\cite{nickels2024documenters,kelly2025bridging}. Within the meeting itself, CommunityClick lets town hall attendees give real-time feedback that is then reported to organizers~\cite{jasim2021communityclick}.

Summarization tools such as MeetingBank reflect a broader effort to build
\genai into civic technology. Large language models have been used to
summarize and cluster large volumes of public input on deliberation platforms
such as Polis~\cite{small2023polis}, and to mediate small-group deliberation,
where model-generated group statements drew broader agreement than statements
written by human mediators~\cite{tessler2024habermas}. Cities have also
deployed LLM-based chatbots for public services, although residents report
difficulty interpreting their responses and understanding how they
work~\cite{zhou2025chatbot}. These uses raise concerns about accuracy: in an
audit of ChatGPT's answers about the 2024 U.S. elections shortly before
election day, about one in ten verifiable claims was inaccurate~\cite{jang2026inaccuracies}. Public
attitudes toward \genai in related information settings are cautious. Across
28 markets, only 36\% of respondents were comfortable with news made mostly by
humans with the help of AI, and only 19\% with news made mostly by AI under
human oversight. Comfort with \genai content was lower for political topics, and most respondents
wanted a human to remain in the loop and AI use to be
disclosed~\cite{rossarguedas2024attitudes}. Far less is known about how
residents view \genai tools for local civic information: whether they are open
to using them, what they would want such tools to provide, and what concerns
would lead them to reject them.

Existing council meeting tools have two gaps. Search tools return matching transcript passages without summarizing them~\cite{brown2021councildataproject,barari2023localview}, and \genai-generated summaries condense these records~\cite{hu2023meetingbank} but, given the accuracy concerns above, are hard to verify. \ourtool{} (Section~\ref{sec:voicesview}) addresses both: it summarizes the public comments made about an issue across meetings and links each statement in the summary to the comments that support it. Users can thus see an overview of public input and check any claim against the original comments, supporting both transparency into local politics and awareness of community perspectives.

\section{Our Tool:  \ourtool{}}\label{sec:voicesview}

\ourtool{} is a tool built to allow users to search and read about issues discussed in city council meetings.\footnote{\ourtool{} is available for Ann Arbor (\url{https://voicesview.fly.dev/?t=VV-VZ6F-XJVE}) and Seattle (\url{https://voicesview-seattle.fly.dev/?t=VV-VZ6F-XJVE}).} \ourtool{} automatically retrieves relevant comments based on local political issues and directly presents them to users, along with a summary of the comments. Provided in the summaries are citations to the public comments for users to read, grounding the \genai-generated summary in actual comments made. \ourtool{} also allows users to watch the video clip of each comment by indexing the video playback at the timestamp associated with the comment. The goal is to create an accessible and transparent tool for users to learn more about what is being discussed in their local council meetings and to trace claims made by the \genai model back to primary sources.

\begin{figure}[t]
  \centering

    \includegraphics[width=.68\linewidth]{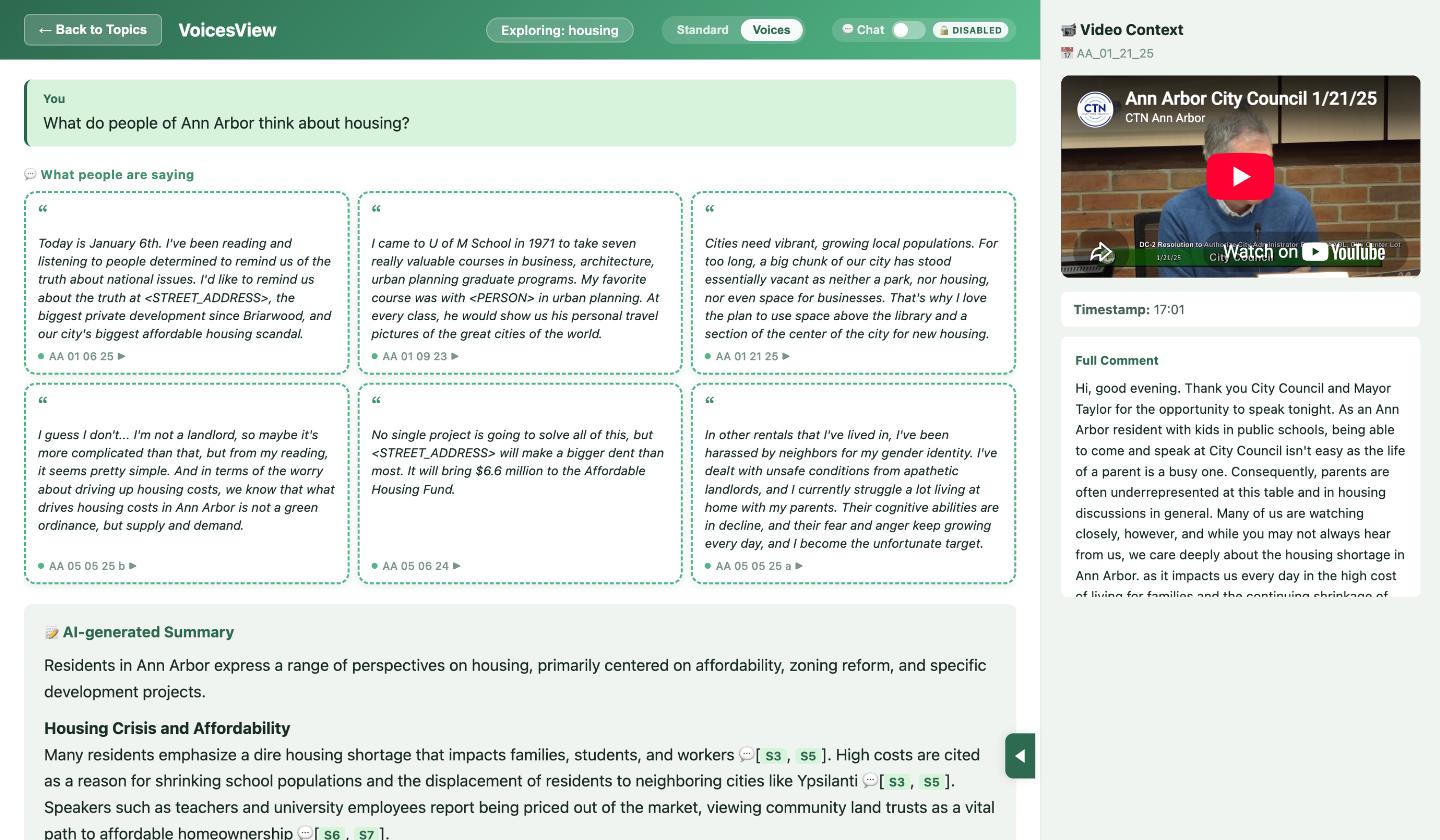}
  \caption{\ourtool{}: A conversation page showing retrieved public comments, an
  AI-generated summary with citations to those comments, and the video
  cued to a selected comment.}
  \Description{A screenshot of the VoicesView conversation page for housing: the user's question at the top, a grid of six quoted resident comments from council meetings, and below them the beginning of an AI-generated summary with bracketed source citations. A side panel on the right shows an embedded YouTube recording of the January 21, 2025 Ann Arbor City Council meeting, the timestamp of the selected comment, and the comment's full text.}
  \label{fig:voicesview}
\end{figure}
Figure~\ref{fig:voicesview} shows how
\ourtool{} accomplishes this with six components.
First, \ourtool{} allows users to either choose from a pre-selected list of issues, or conduct a keyword search for their own issues (see Appendix~\ref{app:homepage} for the home page). Second, \ourtool{} retrieves 10 comments that are deemed relevant to the issue. Third, the response opens with six of the retrieved comments (out of 10) under the heading ``What
people are saying,'' each card carrying an 
excerpt of the one to three consecutive transcribed sentences most relevant to the query. Fourth, appearing below the comment cards is an AI-generated summary of the retrieved comments. Fifth, each retrieved comment can be viewed by the user, both the text transcription of the comment and the video recording. Sixth, every claim made in the AI-generated summary is traceable: selecting a citation in the summary, or a comment card, opens a panel with the meeting recording indexed at the comment timestamp, alongside the comment's full text. 

\paragraph{Curated topic sets.}
The home screen offers four suggested issues: housing; public safety; public services, works, and spaces; and transportation. These four issues were chosen to span the substantive concerns that recur in both cities. They also reflect core responsibilities of US city governments, including policing, housing and development, and the provision of public services~\cite{anzia2022local}, policy areas that differ substantially from the issues that dominate national politics~\cite{anzia2021party}. For these issues, the research team curated the comments (see Appendix~\ref{app:pipeline}) and the summary was generated in advance, so every participant who selected one saw the same content. Issues that participants typed themselves were handled on demand: the retrieval model returns ten comments, which are passed to an enterprise instance of Gemini 3 Flash Preview\footnote{The model is accessed through a
university-managed enterprise agreement under which submitted data remain
private and secure. The agreement approves the service for use with
high-sensitivity data, including electronic Protected Health Information
(ePHI) subject to HIPAA safeguards.} that summarizes them with inline citations to each comment.

\subsection{Data}
 For both Ann Arbor and Seattle,
we collect the recordings of regular city council meetings that the city
publishes online under Sunshine Law provisions: 54 meetings held between January
2023 and December 2025 for Ann Arbor, and 56 meetings held between January 2024
and November 2025 for Seattle. Both corpora comprise regular council
sessions that include the public comment periods -- work and special sessions are excluded. Recordings of each meeting are transcribed with an automatic speech
recognition model, diarized, and segmented into speaker turns, and each turn is
labeled with the meeting section in which it occurs, following the work
of PublicSpeak~\cite{xu2025publicspeak}. Turns located in the public-comment portion
of a meeting are the unit \ourtool{} indexes and presents; we refer to them as
\emph{comments}. Because a resident's entire allotted speaking slot is
contiguous, a comment corresponds to one person's entire speech in a single meeting rather
than to a sentence or a fixed-length passage.

Table~\ref{tab:corpus} summarizes the two corpora. Retrieval is restricted to public comments of at least 200
characters. The restriction to public comments keeps the \genai-generated summaries strictly grounded
in comments made by public commenters, excluding comments made by council members. The length minimum of 200 characters discards procedural comments (e.g., ``Can you hear me now?''). This leaves 647 (Ann Arbor) and 1,292 (Seattle) retrievable comments.
Comments are longer in Ann Arbor (330 words on average, against 163 in
Seattle), likely because Ann Arbor allots 3 minutes per speaker, whereas Seattle allots 2 minutes per speaker. The two datasets contain comparable amounts of public-comment
speech (26.9 and 24.5 hours).

\begin{table}[t]
  \footnotesize
  \setlength{\tabcolsep}{4pt}
  \renewcommand{\arraystretch}{0.95}
  \setlength{\aboverulesep}{0.2ex}
  \setlength{\belowrulesep}{0.3ex}
  \begin{tabular}{@{}lrr@{}}
    \toprule
     & \textbf{Ann Arbor} & \textbf{Seattle} \\
    \midrule
    Council meetings                    & 54 & 56 \\
    Period covered                      & Jan 2023--Dec 2025 & Jan 2024--Nov 2025 \\
    Recording duration (hours)          & 125.0 & 76.6 \\
    Transcript segments                 & 7,981 & 9,965 \\
    Public comments                     & 734 & 1,484 \\
    \quad indexed for retrieval         & 647 & 1,292 \\
    Public-comment speech (hours)       & 26.9 & 24.5 \\
    Mean comment length (words)         & 330 & 163 \\
    \bottomrule
  \end{tabular}
  \caption{Council meeting corpora used by \ourtool{}. Retrieval uses public comments of at
  least 200 characters (Section~\ref{sec:ragdesign}).}
  \label{tab:corpus}
\end{table}

\subsection{RAG Design}
\label{sec:ragdesign}

In this section, we describe how \ourtool{} uses retrieval-augmented generation (RAG) to accurately answer user queries by grounding them in public comments, while using anonymization to reduce risk to commenters.
For each query, \ourtool{} retrieves the ten most relevant public comments and summarizes them. Because short queries such as ``housing'' tend to match comments only superficially related to that topic due to distribution mismatch between user-typed queries and the actual comments, we retrieve with hypothetical document embeddings (HyDE)~\cite{gao2023precise}: a language model drafts an example public comment on the queried issue, and that comment's embedding is used for search. In our evaluation, this raised the precision of retrieved comments from 49.0\% to 84.4\% (Appendix~\ref{app:civicrag}). Before comments reach the user or the language model, the names, street addresses, phone numbers, and email addresses of private individuals are redacted. The ten comments are then summarized by Gemini 3 Flash Preview under instructions to support every claim with an inline citation, describe disagreements among comments, and never invent dates, votes, names, or outcomes. The summarization contains citations which link to supporting comments  (see Appendix~\ref{app:pipeline}).

\section{Study Design}
\label{sec:design}
We conduct a mixed-methods study where we  collect both responses to a novel survey we design and interaction data with our \genai{} civic tech tool (\ourtool). The study took place in Fall 2026.
The target population consists of adults aged 18 or older who live in Ann Arbor or Seattle. We selected these cities because they are geographically diverse, represent two differently sized local governments, and take advantage of our contextual knowledge.

\paragraph{Recruitment and Compensation.}
We recruited participants in both cities through three channels: printed flyers posted in public libraries, advertisements in local publications (\textit{The Stranger} in Seattle and the \textit{Ann Arbor Observer} in Ann Arbor), and targeted advertisements through Meta Ads. In Seattle, where the population was sufficiently large, we also recruited through a survey company (Verasight). Posters were listed in all public library locations in service for both Ann Arbor and Seattle (5 libraries in Ann Arbor and 24 in Seattle). Recruitment materials described the study as research on how people think about local politics. They also stated that the study was not affiliated with any political party, campaign, or candidate, that participation was voluntary, and that participants could skip any question or withdraw at any time. The total time allotted for the study was 75 minutes for the pre-survey, \ourtool{} use, and post-survey. Participants received \$20 for completing both surveys. The study was deemed exempt by institutional IRB.

\paragraph{Eligibility and screening.} The advertisements directed individuals to an online interest form. To be eligible, individuals had to (1) be at least 18 years old, (2) live in Ann Arbor or Seattle, and (3) be able to join a one-hour video session. Residency was verified through the self-provided address information that was required for compensation. 
After noticing anomalies, we implemented several mechanisms to detect non-compliant activity and removed flagged participants from the data analysis (Appendix~\ref{app:fraud}).

\subsection{Procedure}
\label{sec:procedure}

There are three components to this study: (1) pre-survey, (2) use of \ourtool{}, and (3) post-survey. The participants take the pre-survey through Qualtrics and are assigned a personalized code for \ourtool{}. The participant then uses \ourtool{} with their personalized token after completing the pre-survey. Two days after using \ourtool{}, they are sent a post-survey to complete within a week.

\paragraph{Initial survey.} Participants completed an online survey covering demographics and political identification, knowledge of public comment in their city, perceptions of people who speak during public comment, perceptions of local government, community experience and civic engagement, AI literacy and familiarity, issue priorities, and local news sources and information needs. The full survey battery is documented in Appendix~\ref{app:survey}.

\paragraph{Tool instructions and use.} After the survey, participants received instructions on how to use \ourtool{} (Figure~\ref{fig:voicesview}). They were then instructed to use the tool with a suggested duration of at least 10--15 minutes, though they were allotted up to 40 minutes if needed. Participants could select any of the suggested issues or enter issues of their own, and follow-up chat was disabled. We recorded logs of participants' interactions with the tool, stored under a de-identified participant ID. Participants were instructed that the use of \ourtool{} is an open-ended exploration with no objective metric or instructions to follow; they were asked to explore \ourtool{} however they would like. Due to concerns of possible non-compliant submissions, defined as those who partake in the study without good-faith effort, we instructed participants that completion is contingent on substantive use.

\begin{figure*}[t]
  \centering
  \includegraphics[width=0.72\linewidth]{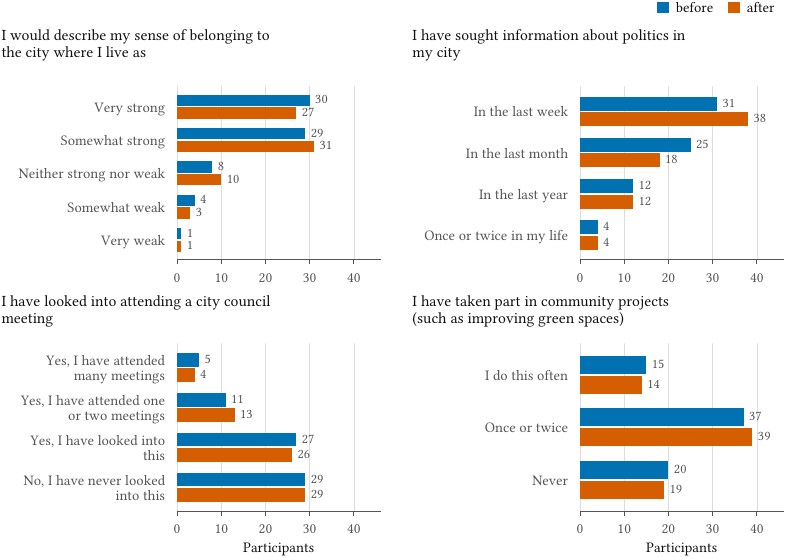}
  \caption{Responses to the four civic-engagement items about \textbf{social capital and civic engagement} before (blue) and after
  (orange) using \ourtool{} ($n = 72$).}
  \Description{Four panels of paired horizontal bar charts: sense of belonging,
  having sought information about city politics, having looked into attending a
  council meeting, and having taken part in community projects. The vertical
  axes give the full response options and the horizontal axes count
  participants. Before bars are above after bars, with counts printed at the
  bar ends. The response distributions are similar across waves.}
  \label{fig:civic}
\end{figure*}

\begin{figure*}[t]
  \centering
  \includegraphics[width=0.72\linewidth]{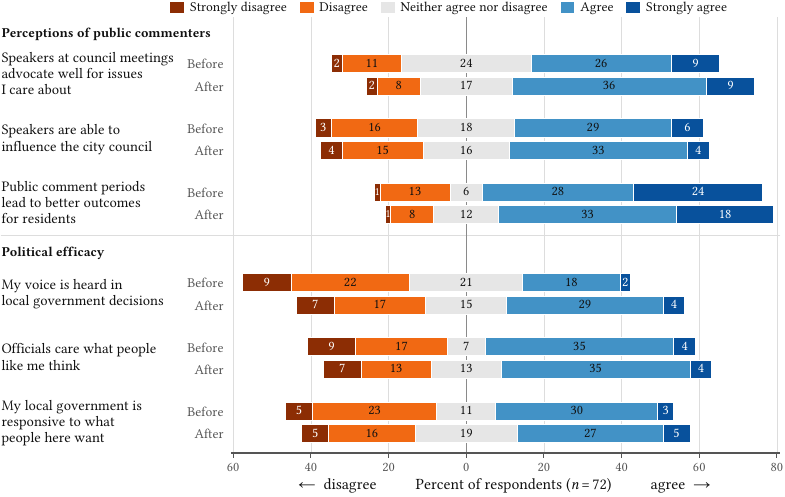}
  \caption{Responses to the agreement items before and after using
  \ourtool{} ($n = 72$), grouped into \textbf{perceptions of public commenters} (top)
  and \textbf{political efficacy} (bottom). Numbers are participant counts.}
  \Description{Six items in two labelled groups, each item with a Before row
  and an After row, drawn as diverging stacked bars from strongly disagree
  through neither to strongly agree. The top group, perceptions of public
  commenters, covers speakers advocating for issues the respondent cares about,
  speakers influencing the city council, and public comment leading to better
  outcomes; the bottom group, political efficacy, covers one's voice being
  heard, officials caring, and local government being responsive. The item
  about one's own voice being heard is the most negative before the session and
  shows the largest shift toward agreement after it; the item about public
  comment leading to better outcomes is the most positive both times; the item
  about speakers influencing the city council barely moves.}
  \label{fig:beliefs}
\end{figure*}

\paragraph{Follow-up survey.} Two days after completing the pre-survey and using \ourtool{}, participants were asked to complete a post-survey. Participants received their compensation after submitting the follow-up survey, as described above. The post-survey was near-identical to the pre-survey, with only demographic questions dropped (e.g., age, educational attainment, political affiliation). 

\section{Results}
\label{sec:results}

We include 72 participants who completed both surveys in the analysis,
24 in Ann Arbor and 48 in Seattle. Forty-three were recruited through flyers,
local publications, and online advertisements, and the other 29 were
Seattle respondents recruited through Verasight. Participants ranged in age from 18 to over 65; 53\% identified as men, 44\%
as women, and 3\% as non-binary or transgender. Most identified as White
(78\%), held a bachelor's degree or higher (72\%), and identified as
Democrats (61\%), with 35\% identifying as independents and 4\% as
Republicans. Nearly two-thirds had lived in their city for more than ten years
(64\%), and 54\% owned their home. As seen in \figref{fig:civic}, they are civically engaged. Full
demographics are in Appendix~\ref{app:demographics}.

Participants drew on several sources of local political information, using
three on average. Most read local or national news (86\%), used social
media (79\%), or talked with others about politics (65\%); fewer watched
local television news (44\%) or talked with neighbors (42\%). Eleven participants used all five sources, and five used only one
(Appendix~\ref{app:news-sources}).

\begin{figure*}[t]
  \centering
  \includegraphics[width=0.725\linewidth]{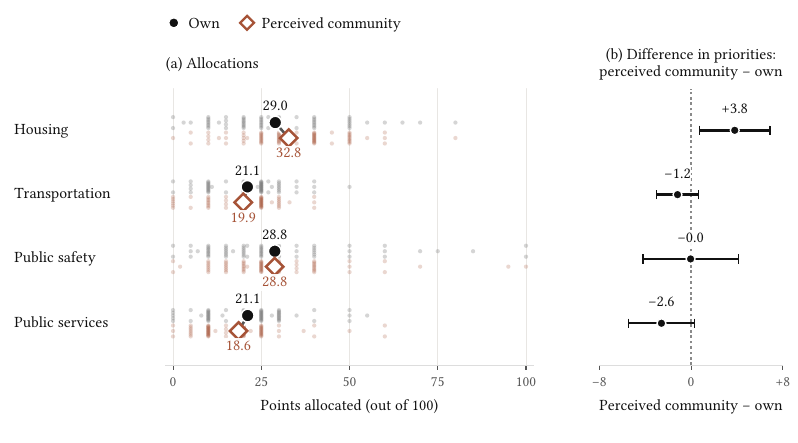}
  \caption{How participants divided 100 points across four issues before using \ourtool{} in the pre-survey, for their own priorities and for what they thought their community prioritizes ($n = 72$). This measures \textbf{how participants perceive alignment with others on local political issues}. (a) Each faint dot is one participant's allocation; large markers show the averages. (b) The average difference between each participant's community and own allocations, with paired 95\% confidence intervals.}
  \Description{Four aligned issue rows. The left panel shows all 72 own and 72 perceived community allocations per issue as faint dots on a 0 to 100 scale. Large filled circles and open diamonds highlight the respective means: housing 29.0 and 32.8, transportation 21.1 and 19.9, public safety 28.8 and 28.8, and public services 21.1 and 18.6 points. The right panel shows paired community-minus-own differences with 95 percent intervals: housing plus 3.8, interval plus 0.7 to plus 6.9; transportation minus 1.2, interval minus 3.0 to plus 0.6; public safety approximately zero, interval minus 4.2 to plus 4.1; public services minus 2.6, interval minus 5.5 to plus 0.3. Only the housing interval excludes zero.}
  \label{fig:allocation}
\end{figure*}

\paragraph{Engagement with \ourtool}
As shown in \figref{fig:engagement} in Appendix~\ref{app:engagement}, \subs{} used \ourtool{} for a median time of approximately 7 minutes. Note that this represents active time, meaning time that was spent using the tool in an active manner (such as opening sources and clicking buttons). We truncate any time spent idle to 120 seconds, which may represent the time a participant spends watching a video recording or reading the summary. Most \subs{} viewed comments in each of the four issues and typed in at least one of their own questions. Housing was the topic participants engaged with most in both cities.

Participants were generally familiar with generative AI. On a five-item true/false quiz about how generative AI works, 49 of the 70 participants who answered got all five items correct (mean 4.6). Most also used AI chatbots regularly: 49 (70\%) used at least one of ChatGPT, Claude, Microsoft Copilot, or Google Gemini several times a week or more, including 23 (33\%) who used one daily, whereas 8 (11\%) had heard of these tools but never used any of them.

\subsection{RQ1: What are existing perceptions of public commenters, perceived alignment with others on local political issues, procedural knowledge about city council meetings, and local political efficacy?}
\label{sec:rq1}

\paragraph{Baseline perception of public commenters} 
In the pre-survey, participants held mixed but generally positive views of
public commenters and public comment (Figure~\ref{fig:beliefs}). About half,
35 (49\%), agreed that public commenters advocate for issues they themselves
care about, 24 were neutral, and 13 disagreed ($M=3.40$)\footnote{Unless noted otherwise, participants rated survey statements on a five-point
scale from 1 (strongly disagree) to 5 (strongly agree), and $M$ denotes the
mean rating across participants. All $p$ values for these comparisons come from
two-sided paired $t$-tests.}. The same number of participants,
35 (49\%), agreed that public commenters can influence the council, while 18
were neutral and 19 disagreed ($M=3.26$). Participants most clearly valued the
role of public comment itself: 52 (72\%) agreed that public comment periods
lead to better outcomes for residents, and only 14 (19\%) disagreed
($M=3.85$).

\paragraph{Perceived Alignment With Others on Local Political Issues}

Here, we ask how participants perceive their community's priorities on local political issues relative to their own priorities. To do so, we tasked participants in the pre-survey with dividing
100 points across four issues: housing, transportation, public safety, and public services. Giving an issue more points means that it matters more to
the respondent. Participants completed this task twice: once for their own priorities, and once for how they believed their community as a whole prioritizes these issues. Prior work gives reason to expect the two allocations may differ. People often believe that their own concerns are shared by few others~\cite{miller1987pluralistic,sparkman2022false}, and they tend to
overestimate how divided the public is~\cite{levendusky2016misperceptions}.
If participants held such misperceptions, their estimates of community
priorities should depart from their own.

Participants gave housing and public safety nearly
equal weight in their own allocations (29.0 and 28.8 points on average), with
transportation and public services each receiving about 21 points. Their
estimates of community priorities followed a similar pattern
(Figure~\ref{fig:allocation}). Housing was the one exception: participants
believed their community prioritized housing more than they did themselves
(32.8 vs.\ 29.0 points; $p = .016$). Differences for the
other three issues were not statistically significant. Overall, participants' own priorities closely tracked the priorities they
attributed to their community. Even on housing, 28 participants gave
themselves and their community the same number of points, 30 gave their
community more, and 14 gave themselves more. Rather than seeing their
concerns as differing from those of people around them, our participants largely
believed that their community shared their priorities.

\paragraph{Procedural Knowledge About Council Meetings}
In the pre-survey, most participants, 53 (74\%), understood that council meetings have a period where members of the public are allowed to speak. However, 15 of those 53 participants did not know what they were allowed to discuss in the public comment period of council meetings. Ann Arbor allows public comment on any topic, whereas Seattle requires comments to be related to agenda items only.

\paragraph{Local Political Efficacy}
Participants' sense of political efficacy -- the belief that they can understand and influence political processes -- was lower and more divided
(Figure~\ref{fig:beliefs}). Only 20 (28\%) agreed that their own voice is
heard in local government decisions, while 31 (43\%) disagreed ($M=2.75$).
Views on whether local government is responsive to what people want were
split, with 33 (46\%) agreeing and 28 (39\%) disagreeing ($M=3.04$). Slightly
more than half, 39 (54\%), agreed that officials care what people like them
think, though 26 (36\%) disagreed ($M=3.11$). Taken together, participants
rated the three efficacy statements lower than the three statements about
public commenters: 49 of the 72 participants rated the public-commenter
statements higher, with an average difference of 0.54 points on the
five-point scale ($p < .001$). In other words, participants
tended to see public comment as a working channel while doubting their own
standing within it and whether comments actually shape local government
decisions.

\subsection{RQ2: How does using a \genai tool impact civic  attitudes and knowledge?}
\label{sec:rq3}

\paragraph{Perceptions of public commenters}

After using \ourtool{}, more participants agreed that public commenters
advocate for issues they themselves care about: 45 (62\%), up from 35 (49\%). Mean
agreement rose from 3.40 to 3.58 ($p=.057$). Twenty participants increased their agreement, 43 gave the same answer, and
nine decreased it (Figure~\ref{fig:beliefs}).

Beliefs about public commenters' influence, by contrast, did not move. Agreement that
public commenters can influence the council went from 35 (49\%) to 37 (51\%), and the
mean remained flat (3.26 to 3.25), with 13 participants moving up and 13 moving down. This result is consistent with our a priori expectation that responses to this statement should not change much in the post-survey. \ourtool{} does not provide information about the voting records associated with these council meeting sessions nor does it provide information on how the public comments were able to influence council members' decisions. Absent this decision-related information, we do not expect \ourtool{} to shift beliefs about the influence of public commenters.

\paragraph{More participants answered the procedure question correctly.}
The share of participants who correctly identified their city's public-comment rule rose from 38 (53\%) to 53 (74\%). Eighteen participants moved from an incorrect to
a correct answer and three moved the other way (exact McNemar $p=.002$). The gain appeared in both cities: from 10 to 17 of 24 in Ann Arbor, and from 28 to 36 of 48 in Seattle.
Of the 18 participants who became correct, 9 had first chosen the other city's rule and the other 9 had answered ``not sure''. Overall, ``not sure'' answers fell from 19 to 8.

\subsection{RQ3: What information would \subs{} like provided in the future by \genai tools?}
\label{sec:rq4}
In the follow-up survey, all 72 participants described what information they
would want from a civic AI tool. We coded these
answers into themes (Figure~\ref{fig:information-needs});
an answer could receive more than one code.

\paragraph{Requests concerned both local issues and council responses.}
The most common request was for information about local issues (23
participants), followed by accurate or well-sourced information (15), council
decisions and follow-through (14), and meeting summaries or records (13).
Fewer asked about how to participate (4) or for comparisons over time or
across cities (4). Several participants
wanted to know whether public input matters; P43 wrote, \emph{``I want to
know if public opinion shaped decisions.''} Others focused on evidence;
P22 asked, \emph{``What sources are credible that people are citing''}.

\paragraph{Some responses raised conditions for use or explicit objections.}
Ten participants raised concerns about AI or set conditions for using it.
Their concerns covered political framing, regulation, the loss of human
interaction, environmental costs, and a preference for traditional reporting.
P26 asked that \emph{``the language is not politically twisted in any way
shape or form.''} P33 wrote, \emph{``It could be useful if it is
regulated,''} and also objected to replacing human interaction and critical
thinking; P63 similarly asked for AI \emph{``being better regulated.''} P34
objected to AI on environmental grounds, and P39 worried that such tools
could replace human conversation and reinforce users' existing views.
\input{figures/facct_v2/figure-information-needs-compact}
\section{Discussion}
\label{sec:discussion}
Our findings suggest that reading actual public comments can shift how residents view public commenters and their own voice in local government.
We do not see a shift
in beliefs about whether public comment influences council decisions, likely because \ourtool{} does not show how council members voted or the reasoning behind their votes. We discuss these findings, participants' perceived alignment with their community, and implications for future \genai civic tools.

\paragraph{Civic attitudes and knowledge}
Before using \ourtool{}, participants valued public comment as an institution but were less certain that the people who speak represent them or can sway the council, and many doubted that their own voice is heard in local government. Although trust in federal institutions has declined nationally, trust in local institutions generally remains high~\cite{hatfield2025issue}; participants' local political efficacy, by contrast, was only moderate and less favorable than their views of public commenters. This suggests that measures of specific attitudes toward local government may capture more variation than general questions about trust. After reading what public commenters had said, participants were more likely to view speakers as advocates for their concerns and to feel that their own voice is heard. 
This is consistent with research on procedural justice, which finds that the opportunity to voice one's views increases perceived fairness even when that voice does not affect the decision~\cite{lind1990voice,tyler2006legitimacy}. Our results suggest that a similar effect may arise when residents see others voice concerns they share. For civic technology, this points to a role that complements local journalism: by making the content of public comment visible, tools like \ourtool{} can show residents that their concerns are being voiced, even if these tools do not show that those concerns are acted upon. Participants appeared to recognize this gap themselves: council decisions and follow-through were among the most common requests for future tools (14 participants; Section~\ref{sec:rq4}). Linking public comments to the votes and actions that follow them may therefore help such tools inform beliefs about influence.

\paragraph{Perceived Alignment With Others on Local Political Issues} 
We find that people generally believe that others in the community place similar priorities on issues as they do. 
The single outlier to that is housing, which \subs{} place the highest priority on, yet believe that others care about more. This is interesting, and warrants further investigation, given the high engagement with this issue in both cities. People may be curious to learn more about what others have to say about this issue \textit{because} they are uncertain of their opinions, or the engagement may just be a reflection of the importance of the issue. Housing is among the most common local concerns raised during public comment in Michigan cities~\cite{ge2025public}, and in Seattle, housing recurs in council discussion~\cite{brown2022councilsinaction} and has been the subject of major land-use reform aimed at affordability~\cite{krimmel2026upzoning}. 
\subs' free-text responses also frequently named housing as the most important issue.

This close alignment between participants' own priorities and those they attributed to their community is surprising, because, a priori, there is little reason to expect the two to align. As local journalism declines, residents increasingly rely on national news, and many remaining local outlets now devote more of their coverage to national politics~\cite{martin2019local,hayes2021newshole}. Because people infer community opinion largely from the news they consume~\cite{mutz1998impersonal}, and national coverage leads people to overestimate how divided the public is~\cite{levendusky2016misperceptions}, we would expect participants to perceive a substantial gap between their own priorities and those of their community. Instead, their own priorities and their estimates of their community's priorities were nearly identical; the only significant difference, for housing, was small (3.8 of 100 points). This pattern contrasts with pluralistic ignorance, in which people believe that their views differ from those of others~\cite{miller1987pluralistic}.

\paragraph{Experience with \ourtool}
Using \ourtool{} generally improved participants' civic attitudes, which is encouraging given that the tool was designed only to provide information. This suggests that baseline attitudes may be under-informed and that exposure to actual council interactions can update them. The effect is most striking for procedural knowledge, which shifted clearly between the pre-survey and post-surveys. Fostering positive experiences and perceptions of local politics, including participation in council meetings, matters because it may encourage engagement and future participation by creating a more conducive space for discussion. 

However, as shown in Appendix~\ref{app:alignment}, \subs{} who did not feel aligned with their community's priorities experienced \ourtool{} differently. Some of their attitudes also became more positive, but generally to a lesser degree, and on some items not at all. One might expect \subs{} who see themselves as outliers in their community to react negatively to seeing public commenters and to become more entrenched in their views. Notably, this did not happen: their perceptions of public commenters as advocates did not become significantly more negative. This gives us more confidence in the tool's potential usefulness for the general public. 

\paragraph{Future design of \genai civic tech}
Participants provided many useful ideas about what kind of features they would like to see built into future tools or iterations of \ourtool. 
These range from information about what the city council is doing to advice on how to best influence its members. 
Overall, \subs{} seem more interested in improved transparency around the activities of city council and the effectiveness of public commenting, than around tracking the public commenters themselves. Participants' second most common request, for accurate and well-sourced information (15 participants), echoes concerns about the accuracy of \genai-generated civic information~\cite{jang2026inaccuracies} and supports designs, like \ourtool{}'s, that link each claim to the exact source it uses.

Ten participants (14\%) raised concerns about \genai, suggesting that skepticism toward \genai civic tools warrants attention in future design. This information should be incorporated into the design process for future tools. If the same quality of information can be provided without the use of AI, designers can likely  attract a broader group of users. 

\section{Limitations}
\label{sec:limitations}

\ourtool{} relies on automatic speech recognition (ASR) to transcribe public comments made at council meetings. Prior research has documented biases in ASR systems, including racial disparities in performance, with substantially higher word error rates for Black speakers than for White speakers~\cite{koenecke2020racial}. Transcription errors may therefore misrepresent what some commenters said, and ASR systems can also hallucinate text that was never spoken, especially in audio with long non-speech segments~\cite{koenecke2024careless}. To mitigate this risk, we took two complementary steps. First, we pre-processed the video recordings to reduce error rates by filtering out segments without any voice activity and using the remaining audio waveform for transcription. Second, the \genai-generated summaries cite the video recording of each comment, so users can verify for themselves what commenters said.

A second limitation is the composition of our sample. Most participants identified as White (78\%), held at least a bachelor's degree
(72\%), and identified as Democrats (61\%). Nearly two-thirds had lived in
their city for more than ten years. We did not selectively target these groups. Rather, they resemble those who are most engaged in politics: education and the resources that accompany it are among the strongest predictors of political participation~\cite{verba1995voice,brady1995beyond}. Those who speak at council meetings are disproportionately longtime residents and homeowners~\cite{einstein2019participates,yoder2020}. Because people are more likely to take part in surveys on topics that interest them, a study centered on local politics likely drew disproportionately from these groups. Future work with larger and more
representative samples could examine whether similar findings occur among
residents who are less educated, less politically engaged, or more conservative.

Relatedly, the study was conducted in two cities, Ann Arbor and Seattle. Both
are highly educated and politically liberal. Council meetings in smaller, more rural, or
more politically divided places may draw different speakers and concerns, and
residents there may respond differently to what they hear. Including a broader
set of localities that varies in size, region, and political composition would
help establish whether these patterns hold more generally.

Finally, apart from one question on public comment procedures, our measures capture how participants perceive local participation, not what they learned about local events. Whether tools such as \ourtool{} can help fill information gaps in local information ecosystems depends on whether people gain lasting knowledge of local events. Measuring this information gain directly, and over a longer period, is a promising direction for future work.

\section{Conclusion}
\label{sec:conclusion}

Prior work has documented who speaks at public meetings; we asked how other
people perceive these speakers, and whether seeing what speakers actually say
changes those perceptions. We built \ourtool{}, a tool that retrieves the
comments members of the public have made at council meetings on a given issue
and summarizes them with citations to the underlying comments, and studied
how 72 participants in Ann Arbor and Seattle responded to it. After using
\ourtool{}, more participants agreed that their own voice is heard in local
government (28\% to 46\%) and that council speakers advocate for issues they
care about (49\% to 62\%), and more correctly identified their city's
public-comment rules. Beliefs that speakers influence the council and that
local government is responsive, by contrast, were essentially unchanged.
In other words, using \ourtool{} increased participants' sense that public
commenters represent their concerns, but not their belief that public comment
influences what the council decides. When asked what they would want from
a civic AI tool, participants most often requested information about local
issues, accurate and well-sourced claims, and records of council decisions,
while ten participants expressed doubts about such tools, citing concerns
about political framing, regulation, and the loss of human interaction. 

As local news coverage declines, residents have fewer ways to learn what
others in their community are saying about local issues. Our findings suggest
that making the record of public comment easier to explore, and grounding AI
summaries of it in what speakers actually said, as participants themselves
requested, can foster more positive
perceptions of local participation.

\clearpage
\section*{Generative AI Usage Statement}

Claude Sonnet 5 was used to assist in writing the code for \ourtool{}. Gemini 3 Flash was used to generate the summaries used in \ourtool{}. Claude Opus 5.5 was used to assist in writing the code that generated the figures. Claude Opus 5.5 was used for editing grammar and refining the clarity of sentences.

\section*{Ethical Considerations Statement}

This study was deemed exempt by our institution's IRB. Participants provided informed consent before beginning the study; recruitment materials stated that the study was not affiliated with any political party, campaign, or candidate, that participation was voluntary, and that participants could skip any question or withdraw at any time. The consent form noted that some comments shown in \ourtool{} discuss political issues that some participants may find sensitive. Survey responses and interaction logs were stored under de-identified participant IDs; names, mailing addresses, and email addresses were collected only to deliver compensation and were kept separate from the analysis data.

During recruitment, we received sign-ups from individuals outside the two study cities. We screened for non-compliant participation using several checks (Appendix~\ref{app:fraud}), including inconsistent addresses across surveys and browser timezones outside the United States. To avoid penalizing eligible residents, no single check led to removal, and flagged participants who completed the study were still compensated. We also required compensation to be delivered physically, which deters participation from outside the study cities.

\ourtool{} displays public comments made by residents who did not consent to this research. Although these comments are public records, \genai tools make them easier to search. We therefore redacted the names, street addresses, phone numbers, and email addresses of private individuals, while retaining those of elected officials, whose statements are a matter of public accountability. Speakers remain identifiable through the linked video recordings. Because \genai summaries can be inaccurate, every claim in a summary links to the comment that supports it, and the model was instructed not to invent dates, votes, names, or outcomes. Queries were processed through an enterprise instance under which submitted data remain private.

\bibliographystyle{ACM-Reference-Format}
\bibliography{sample-base,civicrag,relatedworks}

\clearpage
\appendix
\raggedbottom

\section{Participant demographics}
\label{app:demographics}

Figure~\ref{fig:demographics} reports participants' demographics by city.

\input{figures/facct_v2/figure-demographics}

\clearpage
\section{Local news sources}
\label{app:news-sources}

Figure~\ref{fig:news-upset} shows the combinations of local news sources participants reported using before the study.

\input{figures/facct_v2/figure-news-upset}

\clearpage
\section{Engagement with the tool}
\label{app:engagement}

\begin{figure}[H]
  \centering
  \includegraphics[width=\linewidth]{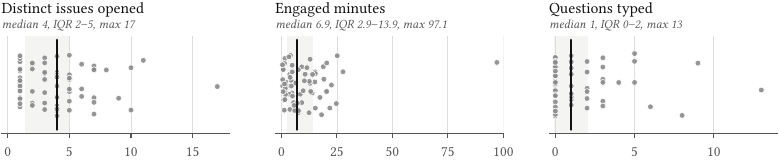}
  \caption{Engagement with \ourtool{} from event logs ($n = 71$; one participant has
  no logs). Each dot is a participant; shaded bands show the interquartile range
  and lines the median. Engaged minutes exclude gaps longer than 120\,s.}
  \Description{Three horizontal strip plots, one per engagement measure. Distinct
  issues opened has a median of 4 and a maximum of 17. Engaged minutes has a
  median of 6.9 and one participant at 97 minutes. Questions typed and answered
  live has a median of 1, with 34 participants at zero. Each panel shows 71
  observations; one participant's missing log is excluded.}
  \label{fig:engagement}
\end{figure}

Table~\ref{tab:logged-events} lists the interaction events the application
recorded. Figure~\ref{fig:engagement} summarizes the three measures reported in
Section~\ref{sec:results}, one mark per participant; the distributions are
long-tailed in every measure, with a single participant accounting for the
97-minute maximum and 34 of the 71 observed participants having no recorded
live question of their own.

\subsection{Engagement with the suggested topics}
\label{app:topic-clicks}

Ann Arbor participants typically opened all four suggested topics, whereas
Seattle participants were more selective, most often opening housing and
public safety (Table~\ref{tab:topic-clicks}).

\begin{table}[H]
  \footnotesize
  \setlength{\tabcolsep}{4pt}
  \begin{tabular}{@{}lrrrrrr@{}}
    \toprule
     & \multicolumn{3}{c}{\textbf{Clicked at least once}}
     & \multicolumn{3}{c}{\textbf{Most-clicked topic}} \\
    \cmidrule(lr){2-4}\cmidrule(l){5-7}
    \textbf{Suggested topic} & Ann Arbor & Seattle & All & Ann Arbor & Seattle & All \\
    \midrule
    Housing          & 22 (92\%) & 32 (67\%) & 54 (75\%) & 5 & 10 & 15 \\
    Public safety    & 20 (83\%) & 26 (54\%) & 46 (64\%) & 2 & 5  & 7  \\
    Public services  & 21 (88\%) & 20 (42\%) & 41 (57\%) & 1 & 2  & 3  \\
    Transportation   & 21 (88\%) & 17 (35\%) & 38 (53\%) & 1 & 5  & 6  \\
    \midrule
    Tied for most clicks &  &  &  & 14 & 23 & 37 \\
    \bottomrule
  \end{tabular}
  \caption{Engagement with the four suggested topics in \ourtool{}, by city
  (Ann Arbor $n = 24$, Seattle $n = 48$). \emph{Clicked at least once}: participants
  who opened the topic from the suggested-topic buttons at least once.
  \emph{Most-clicked topic}: participants whose most frequently opened suggested
  topic was that one; participants who opened two or more topics equally often
  are counted as tied. The buttons appeared in the order listed (housing, public
  safety, public services, transportation) in both cities.}
  \label{tab:topic-clicks}
\end{table}

\begin{table}[H]
  \small
  \setlength{\tabcolsep}{5pt}
  \begin{tabular}{@{}l p{0.68\linewidth}@{}}
    \toprule
    \textbf{Event} & \textbf{What is recorded} \\
    \midrule
    \multicolumn{2}{@{}l}{\emph{Session}} \\
    Session start / end        & Opening and closing the tool, with device context (viewport size, touch input, language, time zone) and the number of issues explored by the end of the session. \\
    Sign-in                    & Successful and failed attempts to enter the participant access code. \\
    Page hidden / visible      & Switching away from and back to the browser tab, used to separate active from idle time. \\
    \midrule
    \multicolumn{2}{@{}l}{\emph{Issue selection}} \\
    Issue input                & Focusing the free-text issue box and beginning to type. \\
    Issue opened               & Selecting an issue, and how: a suggested issue, a typed issue, or a revisit from the list of previously explored issues. \\
    Query completed            & Whether the summary loaded, its latency and length, and the number of comments retrieved. \\
    Issue exited               & Leaving an issue (via the back button or by closing the page), with active and elapsed time on the issue, number of comment and citation clicks, number of videos opened, and maximum scroll depth. \\
    Time limit reached         & The per-issue timer expiring, and the participant dismissing the reminder. \\
    \midrule
    \multicolumn{2}{@{}l}{\emph{Engagement with comments}} \\
    View mode switch           & Toggling between the \emph{Standard} and \emph{Voices} views. \\
    Comment list toggle        & Expanding or collapsing the list of retrieved comments. \\
    Comment / citation click   & Clicking a retrieved comment or an inline citation in the summary, with the comment and meeting IDs and the comment's similarity score; clicks on comments with no available video are logged separately. \\
    \midrule
    \multicolumn{2}{@{}l}{\emph{Video playback}} \\
    Video opened / closed      & Opening a comment's meeting video at its timestamp, and closing it, with time the video was visible and time it was played. \\
    Play / pause / ended       & Playback state changes, with the playback position and the length of each played segment. \\
    Video panel toggle         & Collapsing or expanding the video sidebar. \\
    \midrule
    \multicolumn{2}{@{}l}{\emph{Help}} \\
    About / how-to dialogs     & Opening and closing the \emph{About} and \emph{How to use} dialogs, with time spent in each. \\
    \bottomrule
  \end{tabular}
  \caption{Interaction events logged by \ourtool{}.}
  \label{tab:logged-events}
\end{table}

\subsection{Event sequences within selected visits}

Figure~\ref{fig:session-sequences} complements the all-visit engagement
summaries by retaining the event sequence within each participant's selected
most-engaged visit. Its elapsed-time axis and event-selection rules are
specified in the caption.

\input{figures/facct_v2/figure-session-sequences}

\clearpage
\section{Most important issue facing the city}
\label{app:issue-themes}

\begin{figure}[H]
  \centering
  \includegraphics[width=\linewidth]{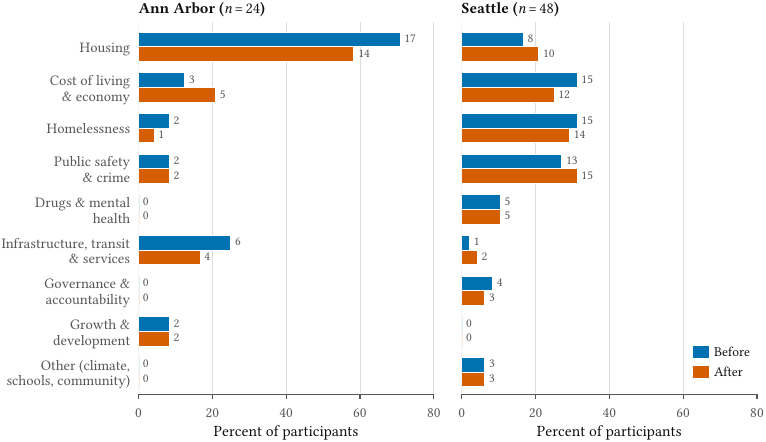}
  \caption{Themes in participants' answers to ``What do you think is the most
  important issue facing your city?'' before and after using \ourtool{}, by
  city ($n = 72$). Bars show the percentage of each city's participants whose
  answer mentioned the theme; numbers are participant counts. An answer could
  mention more than one theme, so percentages do not sum to 100. Two
  post-survey answers that could not be coded are not shown.}
  \Description{Two panels, Ann Arbor with 24 participants and Seattle with 48,
  each with paired before and after horizontal bars for nine themes. In Ann
  Arbor, housing dominates: 17 before and 14 after, followed by
  infrastructure, transit and services (6 and 4) and cost of living (3 and 5).
  In Seattle, answers are spread across cost of living and economy (15 and 12),
  homelessness (15 and 14), public safety and crime (13 and 15), and housing (8
  and 10), with smaller counts for drugs and mental health, governance, and
  other issues. Before and after bars are similar in length for most themes.}
  \label{fig:issue-themes}
\end{figure}

In both surveys, participants answered the open-ended question ``What do you
think is the most important issue facing your city?'' We coded each answer
into one or more themes (Figure~\ref{fig:issue-themes}).

\clearpage
\section{Individual own and perceived-community priorities}
\label{app:allocation-agreement}

Figure~\ref{fig:allocation-agreement} plots each participant's own allocation against their perceived-community allocation for each issue.

\input{figures/facct_v2/figure-allocation-agreement}

\clearpage
\section{Changes in the agreement items}
\label{app:belief-ci}

\begin{figure}[H]
  \centering
  \includegraphics[width=\linewidth]{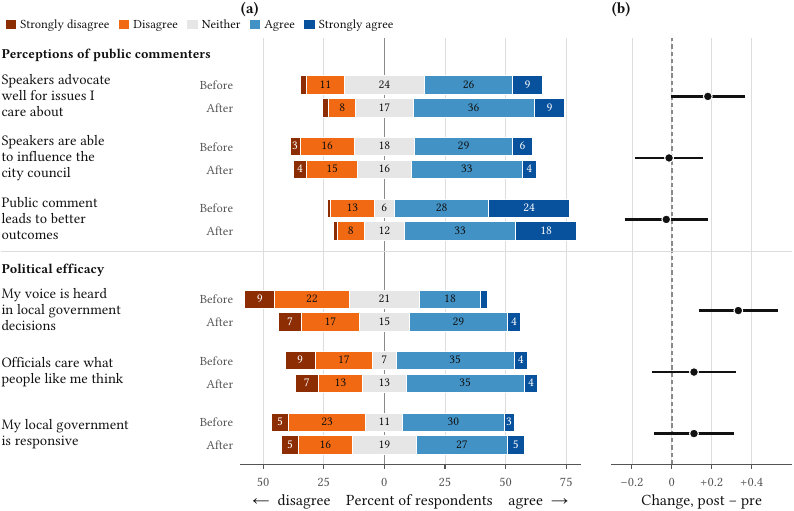}
  \caption{The six agreement items from Figure~\ref{fig:beliefs}, with the
  change in each ($n = 72$). (a) Responses before and after using \ourtool{};
  numbers are participant counts. (b) Mean change from the pre- to the
  post-survey, with paired 95\% confidence intervals. Intervals that do not
  cross the dashed zero line indicate a significant change.}
  \Description{Six statements in two labelled groups, perceptions of public
  commenters and political efficacy. The left panel shows Before and After
  responses as diverging stacked bars. The right panel shows each statement's
  mean change with a 95 percent interval: speakers advocate plus 0.18 (minus
  0.01 to plus 0.37), speakers influence the council minus 0.01, public comment
  leads to better outcomes minus 0.03, voice heard plus 0.33 (plus 0.14 to plus
  0.53), officials care plus 0.11, and local government responsive plus 0.11.
  Only the voice-heard interval lies entirely above zero.}
  \label{fig:belief-ci}
\end{figure}

Figure~\ref{fig:belief-ci} adds the mean pre-to-post change in each of the six agreement items, with paired 95\% confidence intervals, to the response distributions in Figure~\ref{fig:beliefs}.

\clearpage
\section{Paired belief transitions}
\label{app:belief-transitions}

Figure~\ref{fig:belief-transitions} shows how each participant's answer to each agreement item moved between the pre-survey and post-survey.

\input{figures/facct_v2/figure-belief-transitions}

\clearpage
\section{Procedural knowledge by city}
\label{app:knowledge}

\begin{figure}[H]
  \centering
  \includegraphics[width=\linewidth]{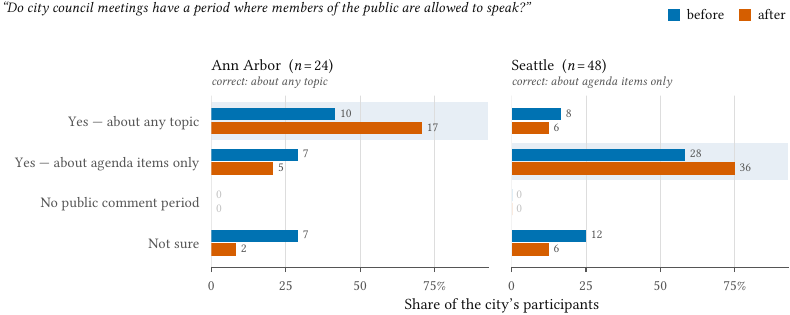}
  \caption{Responses to the procedural-knowledge item before and after using
  \ourtool{}, by city. The correct option, shaded, differs between the two
  cities. Full wording is in Appendix~\ref{app:survey}.}
  \Description{Two panels, Ann Arbor and Seattle, each showing four response
  options with a before bar and an after bar. In Ann Arbor the correct option,
  any topic, rises from 10 of 24 to 17; the other city's rule falls from 7 to 5
  and not sure falls from 7 to 2. In Seattle the correct option, agenda items
  only, rises from 28 of 48 to 36; the other city's rule falls from 8 to 6 and
  not sure falls from 12 to 6. Nobody in either city or wave chose the option
  that there is no public comment period.}
  \label{fig:knowledge}
\end{figure}

Figure~\ref{fig:knowledge} shows responses to the procedural-knowledge item before and after using \ourtool{}, separately for each city.

\clearpage
\section{Perceived alignment and changes in perception}
\label{app:alignment}

\begin{figure}[H]
  \centering
  \includegraphics[width=0.9\linewidth]{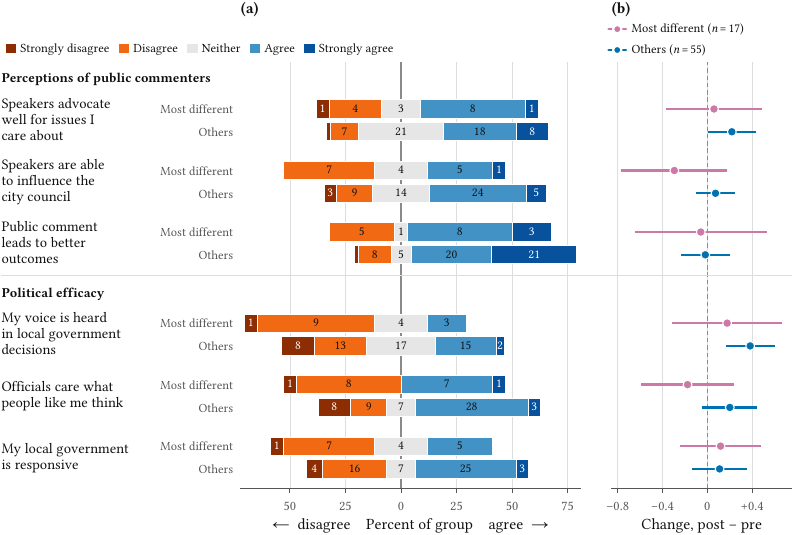}
  \caption{Responses to the agreement items about perceptions of public commenters and political efficacy, \textbf{split by
  perceived alignment: participants whose own priorities differed most from
  those they attributed to their community}, compared with all other
  participants. \emph{Most different}: the 17 participants whose own
  and perceived-community allocations differed by
  more than 20 points, where the difference is the sum of the absolute
  differences across the four issues divided by two (0 = identical, 100 = no
  overlap); \emph{Others}: the remaining 55. (a) Pre-survey responses as a percentage of
  each group; numbers are participant counts. (b) Mean change from the pre- to
  the post-survey, with paired 95\% confidence intervals.}
  \Description{Six statements in two labelled groups, perceptions of public
  commenters and political efficacy, each with a row for the most-different
  group and a row for other participants. The left panel shows pre-survey responses as
  diverging stacked bars from strongly disagree to strongly agree. The
  most-different group is more negative on whether their voice is heard,
  whether local government is responsive, whether speakers can influence the
  council, and whether public comment leads to better outcomes. The right panel
  shows mean pre-to-post changes with 95 percent intervals. Other participants
  gain on voice heard (plus 0.38) and speaker advocacy (plus 0.22); the
  most-different group gains less on these and declines slightly on officials
  caring (minus 0.18) and speakers' influence (minus 0.29). The most-different
  group's intervals are wide and overlap those of other participants.}
  \label{fig:alignment}
\end{figure}

As an exploratory analysis, we examined whether participants who saw their
priorities as differing from their community's responded differently to
\ourtool{}. For each participant, we measured the gap between their own and
perceived-community allocations (Figure~\ref{fig:allocation}) as the number of
the 100 points that would need to move to make the two allocations identical.
We classified the 17 participants whose allocations differed by more than 20
points, roughly the top quarter, as \emph{most different} and compared them
with the remaining 55 (Figure~\ref{fig:alignment}). In the pre-survey, the
most-different group was more negative about whether their voice is heard,
whether local government is responsive, whether speakers can influence the
council, and whether public comment leads to better outcomes
(Figure~\ref{fig:alignment}a). After using \ourtool{}, their gains were also
smaller. Agreement that their voice is heard rose by 0.18 points, compared with
0.38 for other participants, and agreement that speakers advocate for issues
they care about rose by 0.06, compared with 0.22. On whether officials care
what people like them think and whether speakers can influence the council,
the most-different group's ratings declined slightly ($-0.18$ and $-0.29$),
while those of other participants rose (Figure~\ref{fig:alignment}b).
Neither the pre-survey nor the change differences between groups were
statistically significant (Welch's $t$-tests, all $p \geq .11$), likely
because the most-different group is small and therefore underpowered.

\clearpage
\section{Information needs by participant}
\label{app:information-needs}

Figure~\ref{fig:information-needs-full} shows the information-need themes coded for each participant's follow-up answer.

\input{figures/facct_v2/figure-information-needs}

\clearpage
\section{Non-Compliant Participant Screening}
\label{app:fraud}
Some participants joined the Zoom session without having completed the intake form. We asked these participants how they had joined the study and they told us they had received the link from associates. They also admitted to not living in the United States. Given this, we instituted screening for the data used in the analysis. All participants were reimbursed regardless of their suspected location. 

We screened for non-compliant participation using several checks. We flagged participants who (a) reported different addresses in the pre-survey and post-surveys; (b) submitted an obviously false address (e.g., the address of a municipal building); (c) had a browser timezone outside the United States; or (d) had an IP address that resolved to non-residential address space operated by a commercial hosting or VPN provider (e.g., tzulo) rather than to a consumer internet service provider.

Triggering any one of these checks did not automatically lead to removal: participants were removed only when several pieces of evidence together indicated non-compliant participation. After detecting several non-compliant attempts, we also required all compensation to be delivered physically rather than digitally. This served as an additional screen, since participants outside the two study cities would have no way to receive payment.

\clearpage
\input{appendix-survey}

\section{\ourtool{} Home Page}
\label{app:homepage}

\begin{figure}[h]
  \centering
  \includegraphics[width=0.5\linewidth]{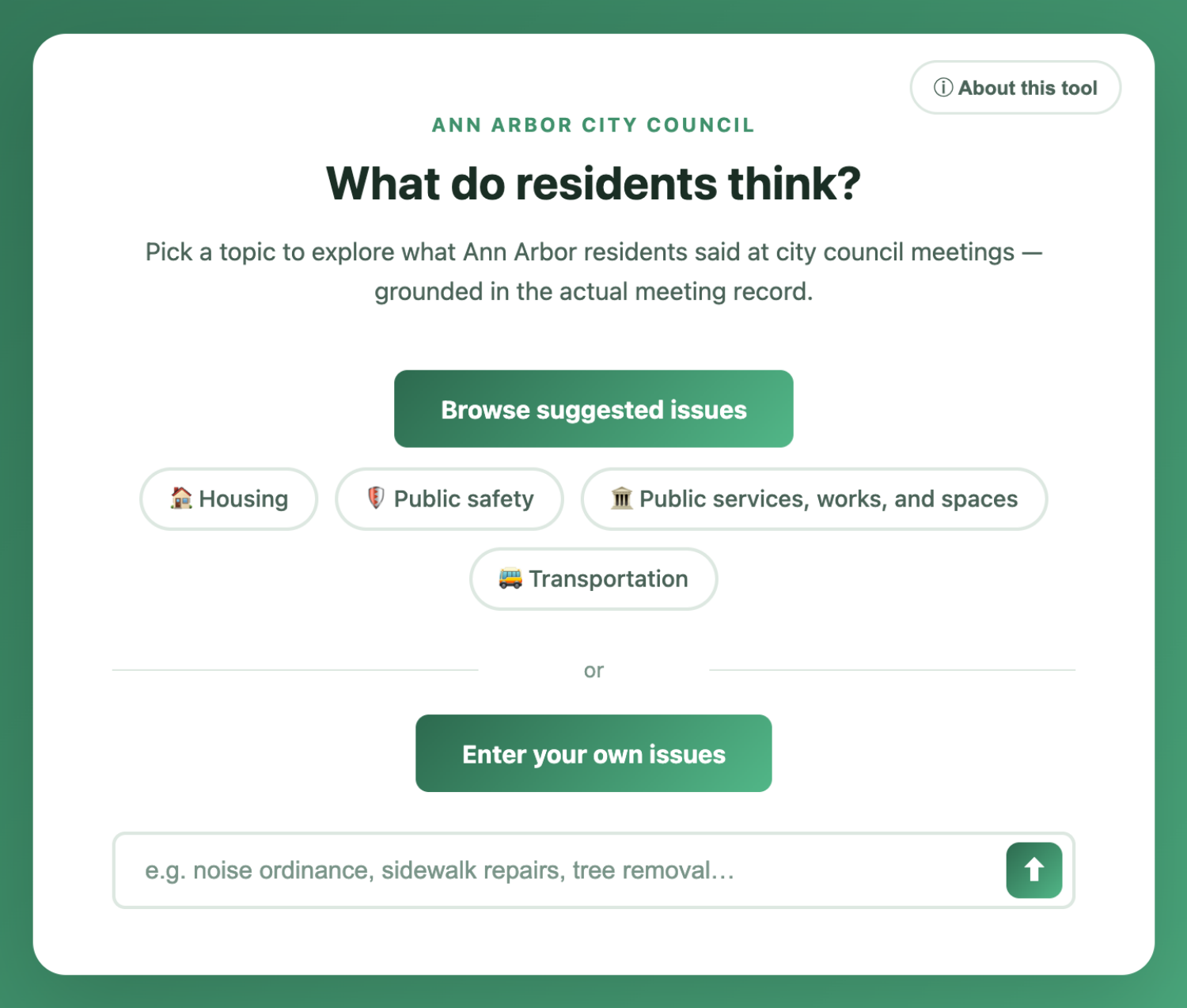}
  \caption{The \ourtool{} home page, where users choose a suggested issue or type their own.}
  \Description{A screenshot of the VoicesView home page, titled ``What do residents think?'', with buttons for suggested issues (housing, public safety, public services, and transportation) and a text box for entering a custom issue.}
  \label{fig:voicesview-home}
\end{figure}

\clearpage
\section{Retrieval and Summarization Pipeline}
\label{app:pipeline}

\begin{figure}[H]
  \centering
  \includegraphics[width=\linewidth]{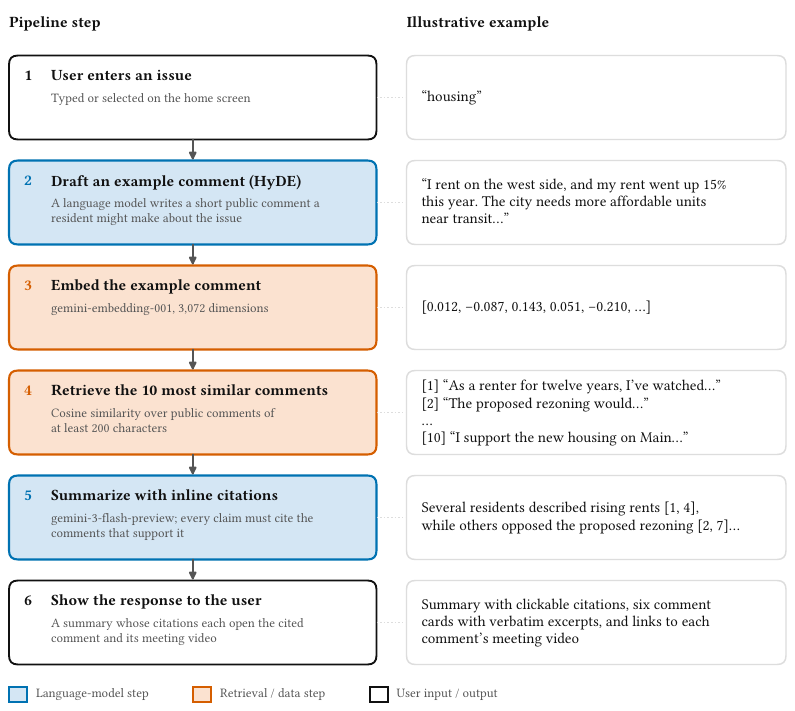}
  \caption{\ourtool{}'s on-demand retrieval and summarization pipeline. Each row is one step; the right column gives an illustrative example of the data at that step (the example comments and summary are invented for illustration). Blue steps use a language model, orange steps are retrieval and data operations, and white steps are the user's input and the response returned to the user. For the four suggested issues, the research team curated the ten comments and the summary was generated in advance (Section~\ref{sec:voicesview}).}
  \Description{A vertical flow diagram with six numbered steps, each paired with an example on the right. Step 1, the user enters an issue, for example "housing". Step 2, a language model drafts an example public comment about the issue (HyDE). Step 3, the example comment is embedded with gemini-embedding-001 into a 3,072-dimensional vector. Step 4, the ten most similar public comments are retrieved by cosine similarity from comments of at least 200 characters. Step 5, gemini-3-flash-preview summarizes the ten comments, citing the comments that support each claim. Step 6, the user sees a summary whose citations each open the cited comment and its meeting video.}
  \label{fig:pipeline}
\end{figure}

This appendix details the pipeline summarized in Section~\ref{sec:ragdesign}. Figure~\ref{fig:pipeline} shows how a query moves through the pipeline, from the issue a user enters to the response \ourtool{} returns.

\paragraph{Indexing and retrieval.}
Each comment is encoded with a general-purpose text embedding model
(Gemini Embedding 001) that produces
3072-dimensional vectors, and is stored in a hosted vector database together with
its meeting identifier and metadata. Retrieval is exact nearest-neighbor search using cosine similarity between the comment and the query. Each query returns $k=10$ comments.

\paragraph{Anonymization.}
Public comments routinely contain
personally identifiable information (PII), including the speaker's name
and home address. While this information is already in a publicly accessible record, \genai-based tools raise privacy concerns by making PII easier to find. To protect commenter privacy, we anonymize comments with Presidio~\cite{presidio}, using spaCy’s en\_core\_web\_sm named entity recognition model and a custom pattern recognizer for street addresses. We replaced each detected span with a
redaction token. Elected officials are exempted through
a list of council members and mayors, which also covers the
misspellings systematically produced by the transcription model, on the grounds
that officials' recorded statements are a matter of public accountability. Note that this process does not completely anonymize public commenters from \ourtool{}: users are still able to watch the video recording to find out who made each comment.

\paragraph{Summarization.}
All ten retrieved comments are numbered and provided to a
language model (Gemini 3 Flash Preview) under a system prompt that
constrains generation in four ways: (1) every factual claim must be supported by a
supplied comment, and insufficient evidence must be reported as such rather than
filled in from parametric knowledge; (2) where the comments conflict, the
disagreement must be described and both sides cited; (3) dates, votes, names, and
outcomes must not be invented; and (4) each claim must carry inline citations
identifying the comments that support it. Citation markers are rendered as hyperlinks that, when clicked, display the underlying comment and the associated video recording.

\paragraph{Curated topic sets.}
For each of the four curated issues, we retrieved 20 candidate comments and reviewed them to assemble a set of ten. We applied four criteria. First, we favored comments for which the issue was the primary focus: given a comment devoted to the issue and one that raised it alongside other topics, or only as a downstream consequence of another issue, we chose the former. Second, we preferred comments that introduced the issue early. Third, we sought balance in stance: when a selected comment took a clear position on the issue, we tried to pair it with a comment taking the opposing position. Fourth, where possible, we drew comments from meetings spread across all time periods in the corpora. 
We then lightly cleaned the selected comments, removing introductory remarks unrelated to the issue (e.g., greetings) and correcting transcription errors. The summary for each issue was generated from exactly that issue's ten cleaned comments, and the resulting comments, excerpts, and summary were cached.

\input{appendix-civicrag-medium}

\end{document}

%% file: figures/facct_v2/figure-information-needs-compact.tex
\begin{figure*}[t]
  \centering
  \includegraphics[width=0.72\linewidth]{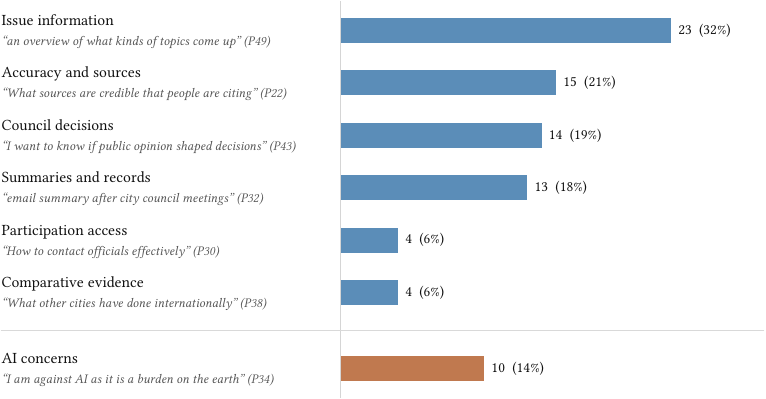}
  \vspace{-3pt}
  \caption{Themes in participants' follow-up answers \textbf{evaluating what a \genai-based tool in civic tech should provide} ($n = 72$), with an example quotation for each. \emph{Issue information}: local issues,
  services, or topics raised at meetings. \emph{Accuracy and sources}: accuracy,
  citations, or credible sources. \emph{Council decisions}: council actions,
  votes, or responses to public input. \emph{Summaries and records}: meeting
  summaries, transcripts, or recordings. \emph{Participation access}: meeting
  dates, sign-ups, or contacting officials. \emph{Comparative evidence}:
  comparisons over time or across cities. \emph{AI concerns}: objections to or
  conditions on using AI.}
  \Description{A horizontal bar chart of information-need themes among 72
  participants. Each theme name has one participant quotation beneath it. Six
  requested types of information are ordered by count: issue information (23),
  accuracy and sources (15), council decisions (14), summaries and records
  (13), participation access (4), and comparative evidence (4). Below a divider,
  in a different color, are AI concerns (10).}
  \label{fig:information-needs}
\vspace{-6pt}
\end{figure*}

%% file: figures/facct_v2/figure-demographics.tex
\begin{figure}[H]
  \centering
  \includegraphics[width=\linewidth]{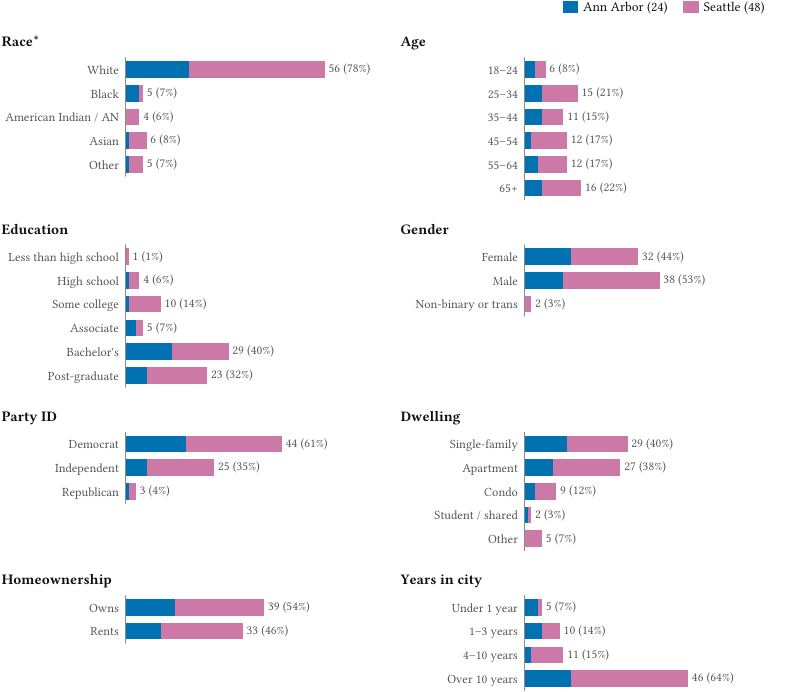}
  \caption{Participant demographics by city ($n = 72$). Labels give the total
  count and its share of the sample. *Race allows multiple selections.}
  \Description{Eight horizontal bar charts of participant demographics, with
  bars stacked by city (24 Ann Arbor, 48 Seattle): race, age, education,
  gender, party identification, dwelling type, homeownership, and years lived
  in the city. Most participants are White (78\%), hold a bachelor's or
  post-graduate degree (72\%), identify as Democrats (61\%), and have lived in
  their city for more than ten years (64\%).}
  \label{fig:demographics}
\end{figure}

%% file: figures/facct_v2/figure-news-upset.tex
\begin{figure}[H]
  \centering
  \includegraphics[width=0.75\linewidth]{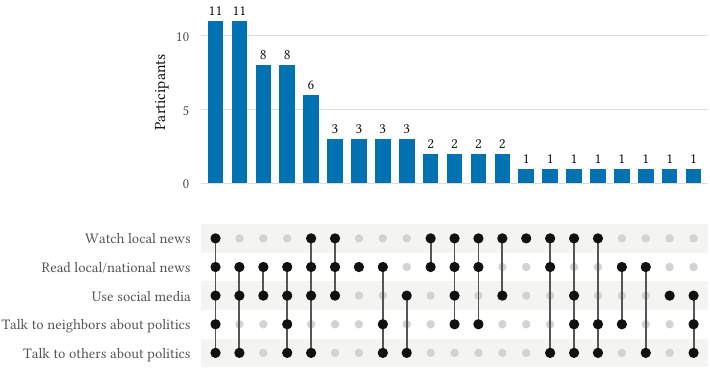}
  \caption{Combinations of local news sources participants reported using before the
  study ($n = 72$). Each column is one combination (black dots); bars count
  participants who selected exactly that combination.}
  \Description{An UpSet plot with five news sources as rows and 21 exact
  combinations as columns. Black dots mark included sources, light dots mark
  excluded sources, and counts appear above the bars. Eleven participants select
  all five sources. Eleven select news reading, social media, and talking to
  others without local television or talking to neighbors. The next two
  combinations each have eight participants.}
  \label{fig:news-upset}
\end{figure}

%% file: figures/facct_v2/figure-session-sequences.tex
\begin{figure}[H]
  \centering
  \includegraphics[width=\linewidth]{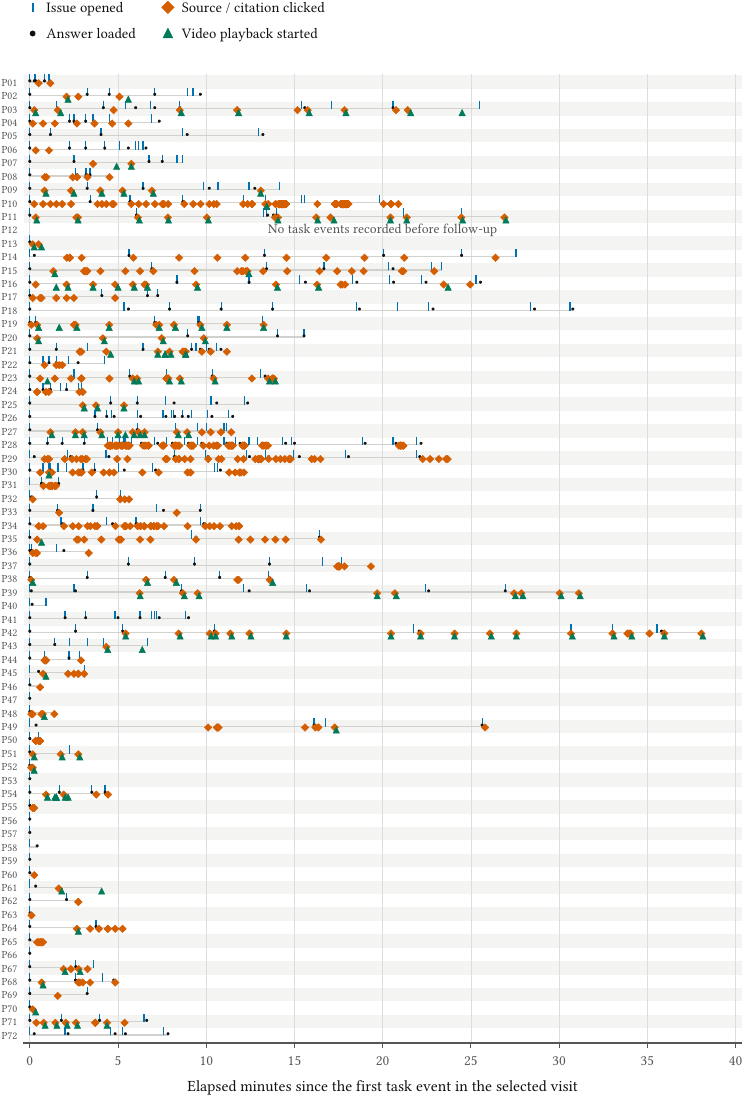}
  \caption{Events in each participant's most-engaged visit before the follow-up
  survey, in minutes from the visit's first event.}
  \Description{An elapsed-time raster with rows P01 through P72 and an axis
  from 0 to 40 minutes. Blue ticks mark opening an issue, black dots an answer
  loading, orange diamonds a source or citation click, and green triangles a
  video playback start. One participant has no logged visit. Selected visits
  span from under a minute to 38 minutes and differ in source and playback
  events.}
  \label{fig:session-sequences}
\end{figure}

%% file: figures/facct_v2/figure-allocation-agreement.tex
\begin{figure}[H]
  \centering
  \includegraphics[width=\linewidth]{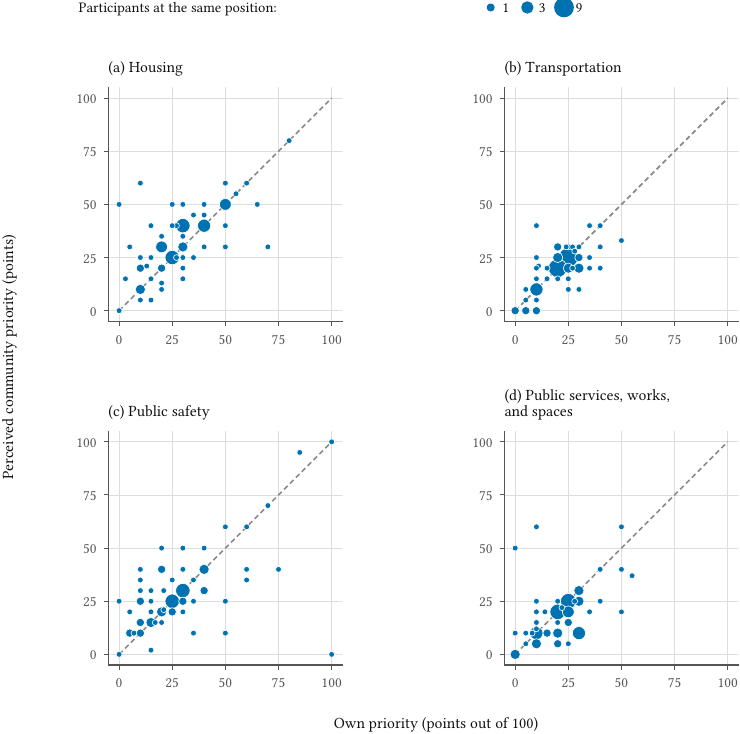}
  \caption{Own allocation (horizontal) against perceived community allocation
  (vertical) for each issue before using \ourtool{} ($n = 72$). Marker area is
  the number of participants at a point; points above the dashed line give the
  community more weight than oneself.}
  \Description{Four square scatterplots for housing, transportation, public
  safety, and public services, works, and spaces. Both axes span 0 to 100 and
  a dashed diagonal marks equality. Marker area represents the number of
  participants at each exact position. Of the 72 participants, the number
  above, on, and below the line is 30, 28, and 14 for housing; 16, 33, and 23
  for transportation; 25, 29, and 18 for public safety; and 12, 28, and 32 for
  public services. All allocations, including extreme
  values of 100, are shown.}
  \label{fig:allocation-agreement}
\end{figure}

%% file: figures/facct_v2/figure-belief-transitions.tex
\begin{figure}[H]
  \centering
  \includegraphics[width=\linewidth]{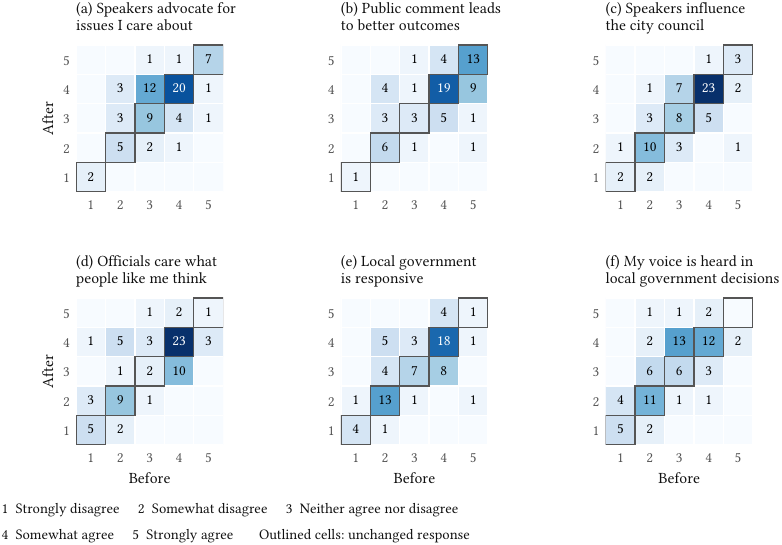}
  \caption{Before (horizontal) and after (vertical) responses on the six agreement
  items, from 1 (strongly disagree) to 5 (strongly agree) ($n = 72$). Cells
  are participant counts; outlined cells are unchanged responses.}
  \Description{Six five-by-five count matrices, one per agreement item. Before
  is horizontal and after is vertical, increasing from 1 to 5. Counts appear
  in occupied cells and unchanged-response cells are outlined. Of the 72
  participants, the number whose response increased, stayed the same, or
  decreased was: speakers advocating for issues participants care about, 20,
  43, and 9; public comment leading to better outcomes, 13, 42, and 17;
  speakers influencing the city council, 13, 46, and 13; officials caring, 16,
  40, and 16; local government being responsive, 17, 43, and 12; and one's own
  voice being heard, 29, 34, and 9.}
  \label{fig:belief-transitions}
\end{figure}

%% file: figures/facct_v2/figure-information-needs.tex
\begin{figure}[H]
  \centering
  \includegraphics[width=0.88\linewidth,height=0.78\textheight,keepaspectratio]{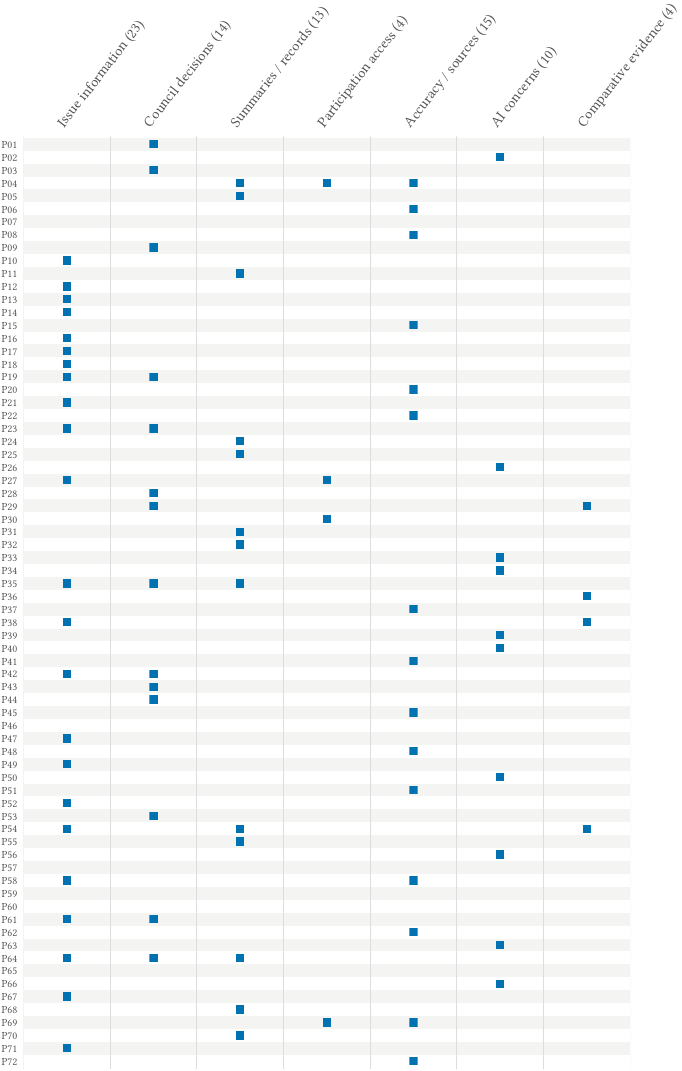}
  \caption{Information-need themes coded for each participant ($n = 72$).
  \emph{Issue information}: local issues, services, or topics raised at
  meetings. \emph{Council decisions}: council actions, votes, or responses to
  public input. \emph{Summaries / records}: meeting summaries, transcripts, or
  recordings. \emph{Participation access}: meeting dates, sign-ups, or
  contacting officials. \emph{Accuracy / sources}: accuracy, citations, or
  credible sources. \emph{AI concerns}: objections to or conditions on using
  AI. \emph{Comparative evidence}: comparisons over time or across cities.}
  \Description{A 72-row matrix with seven columns: issue information (23),
  council decisions (14), summaries or records (13), participation access (4),
  accuracy or sources (15), explicit AI concerns (10), and comparative evidence
  (4). Filled squares mark assigned themes.}
  \label{fig:information-needs-full}
\end{figure}

%% file: appendix-survey.tex
\section{Survey Instrument}
\label{app:survey}

Below is the survey administered to participants. Question
wording and response options are reported verbatim. Unless otherwise noted, items are
single-choice questions presented with vertically listed response options.
Horizontal rules separate survey pages. The follow-up survey repeated these
items, except for the demographic and political-identification items
(Q2--Q12). The version of both surveys administered to participants recruited
through Verasight additionally asked participants whether they were accessing
the survey from a phone, a tablet, a computer, or another device.

\providecommand{\surveyopts}[1]{\par\smallskip{\small\textit{Response options:} #1}}
\providecommand{\surveynote}[1]{\par\smallskip{\small\textit{Note:} #1}}
\providecommand{\surveypage}{\par\medskip\noindent\rule{\linewidth}{0.4pt}\par\medskip}

\subsection{Demographics and Political Identification}

\begin{enumerate}
  \item[Q1.] I would describe my age as
  \surveyopts{under 18; 18 -- 24; 25 -- 34; 35 -- 44; 45 -- 54; 55 -- 64; 65 +}

  \item[Q2.] I own (or have an ownership claim to) the property where I live
  \surveyopts{Yes; No}

  \item[Q3.] I would describe my dwelling as
  \surveyopts{Single-family home; Apartment; Condo; Student or shared housing;
  Other (with a text box)}

  \item[Q4.] I am \\ \emph{Select all races that apply}
  \surveyopts{White; Black or African American; American Indian or Alaska
  Native; Asian; Prefer not to answer; Other (with a text box)}
  \surveynote{Multiple selection.}

  \item[Q5.] My Hispanic or Latino origins are \\ \emph{Select all ethnicities
  that apply. If none apply select ``I don't have Hispanic origins''}
  \surveyopts{I don't have Hispanic origins; Mexican; Puerto Rican; Cuban; I
  refer to my Hispanic origins as (with a text box); Prefer not to answer}
  \surveynote{Multiple selection.}

  \item[Q6.] The highest level of school I have completed or the highest degree I
  have received is
  \surveyopts{Less than high school degree; High school graduate (high school
  diploma or equivalent including GED); Some college but no degree; Associate
  degree in college (2-year); Bachelor's degree in college (4-year); Any form of
  post-graduate degree}

  \item[Q7.] I would describe my gender identity as
  \surveyopts{Female; Male; Non-binary or transgender; Prefer not to answer}

  \item[Q8.] My zip code is \\ \emph{Type your zip code below.}
  \surveynote{Open text entry.}

  \item[Q9.] Generally speaking, I usually think of myself as a Republican, a
  Democrat, or an Independent
  \surveyopts{Democrat; Independent; Republican}

  \item[Q10.] Would you call yourself a strong Democrat or not a very strong
  Democrat?
  \surveyopts{Strong Democrat; Not very strong Democrat}
  \surveynote{Shown only to participants who selected ``Democrat'' on Q9.}

  \item[Q11.] Would you call yourself a strong Republican or not a very strong
  Republican
  \surveyopts{Strong Republican; Not very strong Republican}
  \surveynote{Shown only to participants who selected ``Republican'' on Q9.}

  \item[Q12.] Do you think of yourself as closer to the Republican Party or the
  Democratic Party?
  \surveyopts{Closer Democratic; Neither; Closer Republican}
  \surveynote{Shown only to participants who selected ``Independent'' on Q9.}

  \item[Q13.] I am based in
  \surveyopts{Ann Arbor; Seattle}
\end{enumerate}

\surveypage

\subsection{Knowledge of Public Comment}

\noindent First, we are going to ask you about city council meetings in your city.

\begin{enumerate}
  \item[Q14.] Do city council meetings have a period space where members of the
  public are allowed to speak?
  \surveyopts{Yes, city council meetings in my city have a period where the
  public can speak about items on the meeting's agenda; Yes, city council
  meetings in my city have a period where the public can speak about anything;
  No, city council meetings in my city do not provide any opportunity for the
  public to speak; I'm not sure}
\end{enumerate}

\surveypage

\noindent After Q14, participants saw feedback describing public comment in
their city, preceded by ``That's right!'' when their answer was correct:
\begin{itemize}
  \item \emph{Seattle:} ``(That's right!) In Seattle, people are invited to speak
  about agenda items during the public comment period.''
  \item \emph{Ann Arbor:} ``(That's right!) In Ann Arbor people are invited to
  speak about any topic during the public comment period.''
\end{itemize}
Each version continued: ``Now we have a few questions to learn what you think
about the people who generally speak during the public comment period.''

\subsection{Perceptions of Public Commenters}

\noindent Each statement was followed by the instruction \emph{Please indicate
how much you agree or disagree with the statement above}, with the response
options: Strongly agree; Somewhat agree; Neither agree nor disagree; Somewhat
disagree; Strongly disagree.

\begin{enumerate}
  \item[Q15.] I trust that people who typically speak at city council meetings in
  my city are \textbf{\uline{good advocates for the issues that I care about}}

  \item[Q16.] I believe that public comment periods \textbf{\uline{lead to better
  outcomes}} for the residents of my city

  \item[Q17.] I believe that people who typically speak at city council meetings
  in my city are \textbf{\uline{able to influence}} the city coucil
\end{enumerate}

\surveypage

\subsection{Perceptions of Local Government}

\noindent Now we have some questions to learn what you think about your local
government.

\noindent Each statement used the same instruction and five-point agreement scale
as Q15--Q17.

\begin{enumerate}
  \item[Q18.] Most officials in my local government \textbf{\uline{care what people
  like me think}}

  \item[Q19.] My local government is \textbf{\uline{responsive to what the people in
  my area want}}

  \item[Q20.] I feel that \textbf{\uline{my voice is heard}} in local government
  decisions that affect me
\end{enumerate}

\surveypage

\subsection{Community Experience and Civic Engagement}

\noindent Now we have some questions to learn about your experience in your city.

\begin{enumerate}
  \item[Q21.] I have lived in this city
  \surveyopts{Less than 1 year; 1-3 years; 4-10 years; More than 10 years}

  \item[Q22.] I would describe my sense of belonging to the city where I live as
  \surveyopts{Very strong; Somewhat strong; Neither strong nor weak; Somewhat
  weak; Very weak}

  \item[Q23.] I have taken part in community projects (such as improving green
  spaces)
  \surveyopts{I do this often; Once or twice; Never}
\end{enumerate}

\surveypage

\begin{enumerate}
  \item[Q24.] I have sought information about politics in my city
  \surveyopts{Once or twice in my life; In the last year; In the last month; In
  the last week}

  \item[Q25.] I have looked into attending a city council meeting
  \surveyopts{No, I have never looked into this; Yes, I have looked into this;
  Yes, I have attended one or two meetings; Yes, I have attended many meetings}
\end{enumerate}

\surveypage

\subsection{AI Literacy and Familiarity}

\noindent Next, we have some questions about artificial intelligence (AI). An AI
system is a computer-based system capable of making recommendations, predictions,
or decision. For example, prominent examples of generative AI are ChatGPT and
Copilot.

\begin{enumerate}
  \item[Q26.] Within the larger field of Artificial Intelligence (AI), generative
  AI is a particular type of AI that produces original content based on patterns
  and examples it has learned from a large amount of existing data.
  Examples of generative AI include: ChatGPT, Copilot, and Gemini.
  \begin{enumerate}
    \item[(a)] When generative AIs (like ChatGPT) answer you, they calculate the
    probability of the next words one after another to form sentences
    \item[(b)] Generative AIs (like ChatGPT) are based only on sources that are
    trustworthy and knowledgeable in the topic
    \item[(c)] When using generative AI, the phrasing of the prompt has no effect on
    the output
    \item[(d)] Sometimes, users receive different answers from a generative AI, even
    though they asked the same question
    \item[(e)] Sometimes it is difficult to detect whether a generative AI is
    providing wrong information
  \end{enumerate}
  \surveyopts{True; False (matrix table, one row per statement)}

  \item[Q27.] Have you ever heard of or used the following technologies?
  \begin{enumerate}
    \item[(a)] ChatGPT
    \item[(b)] Claude
    \item[(c)] Microsoft Copilot
    \item[(d)] Google Gemini
  \end{enumerate}
  \surveyopts{I'm hearing about it here for the first time; I heard the name, but
  never used it; I used it once or twice before; I use it several times a week; I
  use it daily (matrix table, one row per technology)}
\end{enumerate}

\surveypage

\subsection{Issue Priorities}

\begin{enumerate}
  \item[Q28.] \emph{Imagine you have 100 points and you want to
  distribute these such that issues you care about have more points and issues
  you don't care about have fewer points.} \\ You are given 100 points. Please place these points on the
  following issues. You should place more points on issues that you think are
  more important \textbf{\uline{to you}} and place fewer points on the issues
  that you think are less important \textbf{\uline{to you.}}
  \surveyopts{Housing; Transportation; Public Safety; Public services, works, and
  spaces}
  \surveynote{Constant-sum question: participants entered a number of points for
  each issue, and a running total was displayed. Allocations had to sum to 100.}

  \item[Q29.] \emph{Imagine you have 100 points and you want to
  distribute these such that issues you care about have more points and issues
  you don't care about have fewer points.} \\ You are given 100 points. Please place these points on the
  following issues. You should place more points on issues that you think are
  more important for people in \textbf{\uline{your city}} and place fewer
  points on the issues that you think are less important to the people in
  \textbf{\uline{your city.}}
  \surveyopts{Housing; Transportation; Public Safety; Public services, works, and
  spaces}
  \surveynote{Constant-sum question, as in Q28.}

  \item[Q30.] What do \textbf{\uline{you think}} is the most important issue
  facing your city?
  \surveynote{Open text entry.}

  \item[Q31.] What do you think \textbf{\uline{others in your city}} think is
  the most important issue facing your city?
  \surveynote{Open text entry.}
\end{enumerate}

\surveypage

\subsection{Local News Sources and Information Needs}

\begin{enumerate}
  \item[Q32.] I get news about local politics from these sources at least once a
  month \\ \emph{Please select all of the sources below where you learn about
  local politics regularly}
  \surveyopts{Watch local news; Read local/national news; Use social media; Talk
  to neighbors about politics; Talk to others about politics; None of these}
  \surveynote{Multiple selection.}

  \item[Q33.] Given access to a chatbot service which would answer questions
  about the types of things that public talks about in city council meetings I
  would want to know \\ \emph{Please let us know what you think would be useful
  information for a future AI tool to provide}
  \surveynote{Open text entry.}
\end{enumerate}

\surveypage

\noindent The survey closed with the message: ``We thank you for your time spent
taking this survey. Your response has been recorded.''

%% file: appendix-civicrag-medium.tex
\section{Empirical Validation of the Retrieval System}
\label{app:civicrag}

The AI tool used in this study is powered by \textsc{CivicRAG}, a
retrieval-augmented generation pipeline that answers questions about a
community by retrieving relevant public comments from recorded city council
meetings and presenting them -- together with the source video segments -- under a
short generated summary. Because participants use the tool to form impressions of
what community members say in these meetings, the study's validity depends on the
retrieval layer surfacing comments that genuinely bear on a user's question. This
appendix reports a corpus-grounded evaluation of that layer, covering the data,
the verifiable evaluation design, the metrics, headline findings, and the full
set of retrieval and prompt configurations we compared.

\subsection{Data}
The evaluation corpus comprises the public-comment portions of the [LOCALITY]
city council meetings held during 2023. Meeting recordings were transcribed with
an automatic speech recognition model~\cite{xu2025publicspeak} and segmented into
individual comments, yielding a corpus of 912 comments.

\subsection{Evaluation design}
A central obstacle to evaluating a civic RAG system is the absence of ground
truth: a question such as ``What issues do residents care about?'' has no
objectively verifiable answer. We therefore restrict evaluation to queries whose
answers can be computed directly from the corpus. Each query is instantiated from
a fixed template with a single topic~$t$: \emph{``Do people in [LOCALITY] care
about \{topic\}? Provide: (1)~a prevalence estimate as a percentage; (2)~six
verbatim quoted comments from the documents above that discuss \{topic\}.''} This
elicits two checkable outputs: a \emph{prevalence estimate} $\hat{p}(t)$, the
percentage of comments in the corpus that concern~$t$, and a set of six
\emph{example comments}. Ground-truth prevalence is the global topic frequency in
the corpus; whether a given comment is on-topic is decided by a BERT classifier
fine-tuned on council-meeting comments~\cite{ge2025public}. We evaluate 20 topics
spanning civic domains and social issues (listed at the end of this appendix).

\subsection{Metrics}
We report, averaged over four LLM/retriever configurations (see
\emph{Model configurations} below):
\begin{itemize}
  \item \textbf{MAE}: mean absolute error between the estimated prevalence
  $\hat{p}(t)$ and the corpus ground truth, with prevalence treated as a
  proportion in $[0,1]$ (lower is better).
  \item \textbf{P@6} and \textbf{P@20}: precision of, respectively, the six
  comments passed to the generator and the 20-comment pool from which they are
  drawn (higher is better).
\end{itemize}
Table~\ref{tab:civicrag-scenarios} reports all three across retrieval strategies.

\begin{table}[t]
  \footnotesize
  \setlength{\tabcolsep}{4pt}
  \begin{tabular}{@{}l p{4.5cm} ccc@{}}
    \toprule
    \textbf{Sc.} & \textbf{Description} &
    \textbf{MAE}\,$\downarrow$ & \textbf{P@6}\,$\uparrow$ & \textbf{P@20}\,$\uparrow$ \\
    \midrule
    \multicolumn{5}{@{}l}{\textit{Semantic retrieval baselines}}\\
    S1  & Topic labels {}+{} corpus-level prevalence prior (near-oracle bound for MAE) & \textbf{0.071} & 49.4 & 35.2 \\
    S2  & Topic labels only; no prevalence hint & 0.317 & 54.6 & 35.2 \\
    S3  & Raw comment text only & 0.512 & 49.0 & 35.2 \\
    \addlinespace
    \multicolumn{5}{@{}l}{\textit{Random retrieval ablations}}\\
    S4  & Topic labels; comments sampled uniformly at random & 0.091 & 21.3 & 10.5 \\
    S5  & Raw text only; comments sampled at random & 0.278 & 17.3 & 10.5 \\
    S6  & Hybrid: each round samples 20 comments at random, keeps top 2 by similarity & 0.447 & 37.7 & 21.8 \\
    S7  & Random comments with post-hoc similarity scores appended & 0.500 & 18.8 & 10.5 \\
    \addlinespace
    \multicolumn{5}{@{}l}{\textit{Multi-stage approaches}}\\
    S8  & Two-stage, shared random pool: top-$K$ by similarity for examples, full pool for prevalence & 0.262 & 25.4 & 14.6 \\
    S9  & As S8, plus threshold-based prevalence hint $\hat{p}=|\{d:\mathrm{sim}(d,q)>\tau\}|/N$ & 0.199 & 26.5 & 14.6 \\
    S10 & Decoupled pools: semantic top-$K$ from full corpus for examples, separate random sample for prevalence & 0.293 & 50.0 & 35.2 \\
    \addlinespace
    \multicolumn{5}{@{}l}{\textit{Retrieval query reformulation (decoupled, as S10)}}\\
    S12 & Rocchio pseudo-relevance feedback: query vector moved toward pseudo-relevant centroid & 0.196 & 51.2 & 34.8 \\
    S14 & TF-IDF query expansion: top-$M$ terms from an initial pass appended to the query & 0.177 & 55.4 & 34.5 \\
    S15 & HyDE: retrieve with the embedding of an LLM-generated hypothetical comment & 0.220 & \textbf{84.4} & \textbf{67.5} \\
    \bottomrule
  \end{tabular}
  \caption{Scenario comparison across retrieval methods and prompt design,
  averaged over four LLM and embedding configurations
  (Qwen3-\{8B,\,14B\}\,$\times$\,Qwen3-Embedding-\{0.6B,\,8B\}). MAE measures
  prevalence-estimation error (lower is better); P@6 and P@20 measure the
  precision of the six comments delivered to the LLM and of the 20-comment
  retrieval pool (higher is better). Bold marks the best value per column. S1
  includes a corpus-level prevalence prior and is an approximate upper bound for
  MAE.}
  \label{tab:civicrag-scenarios}
\end{table}

\subsection{Findings}
\emph{Retrieval.} Query reformulation markedly improves retrieval quality. The
HyDE pipeline~\cite{gao2023precise} (S15), which retrieves using the embedding of
an LLM-generated hypothetical comment rather than the raw query, reaches
$\text{P@6}=84.4\%$ and $\text{P@20}=67.5\%$, against $49.0\%$ and $35.2\%$ for
standard dense retrieval (S3). We attribute the gain to the reformulated query
spanning more of the vocabulary and context that surround a civic topic in
practice. This suggests that off-the-shelf dense retrieval is a weak default for
social-science tasks, where one concept admits many operationalizations and user
queries are typically short and underspecified; with reformulation, the tool more
reliably surfaces the community voices relevant to a given concern.

\emph{Prevalence estimation.} MAE is comparatively forgiving but exposes a
distinct failure mode. Even in S1, where the true prevalence is placed directly
in the prompt, MAE is $0.071$ rather than $0$: for contested topics such as
policing, the Qwen3-8B model sometimes rejects the supplied value and
substitutes an estimate produced by its own reasoning.

\subsection{Model configurations}
All numbers in Table~\ref{tab:civicrag-scenarios} are averaged over the four
configurations of a $2\times2$ grid crossing two language models
(Qwen3-8B, Qwen3-14B) with two embedding models (Qwen3-Embedding-0.6B,
Qwen3-Embedding-8B).

\subsection{Two-stage design}
Scenarios S8 onward adopt a decoupled pipeline that separates the two outputs.
One LLM call selects the illustrative example comments (Part~1); a second,
independent call estimates topic prevalence from a randomly sampled pool
(Part~2). Decoupling keeps retrieval that is tuned for example quality from
biasing the prevalence estimate, which needs an unbiased sample of the corpus.

\subsection{Scenario descriptions}
Each scenario varies along two axes: how comments are retrieved (the pool shown
to the LLM), and what auxiliary information the prompt carries (topic labels,
prevalence priors, similarity scores).

\paragraph{Semantic retrieval baselines (S1--S3).}
All three rank the full corpus by cosine similarity between the query embedding
and each comment embedding and return the top-$K$; they differ only in auxiliary
information. \textbf{S1} adds per-document topic labels (binary indicators from
the fine-tuned BERT classifier) \emph{and} the true corpus-level prevalence, so
its residual MAE reflects only the LLM's failure to reproduce a value it was
given. \textbf{S2} withholds the prevalence, leaving the LLM to infer it from the
labels on the $K$ retrieved comments. \textbf{S3} withholds both: the LLM sees
only raw comment text and must jointly judge relevance and estimate prevalence.

\paragraph{Random retrieval ablations (S4--S7).}
These replace semantic retrieval with uniform random sampling to isolate the
effect of retrieval quality. \textbf{S4} attaches BERT topic labels to randomly
drawn comments (compare with S2 to isolate retrieval quality when labels are
present). \textbf{S5} provides random comments with no labels (compare with S3 to
isolate the benefit of semantic retrieval alone). \textbf{S6} is an iterative
hybrid: in each of $K$ rounds it samples 20 comments at random and keeps the two
most similar to the query, yielding a pool that mixes broad coverage with
relevance filtering. \textbf{S7} provides random comments each annotated with its
(post-hoc) cosine similarity to the query, testing whether a numeric relevance
signal alone can substitute for relevance-based selection.

\paragraph{Multi-stage approaches (S8--S10).}
\textbf{S8} draws a large random pool of $N$ comments; Part~1 takes the top-20 by
similarity from that pool and asks the LLM to choose six representative quotes,
while Part~2 shows all $N$ comments (with similarity scores) for prevalence
estimation. \textbf{S9} adds to S8's Part~2 a threshold-based prevalence hint,
$\hat{p}=|\{d:\mathrm{sim}(d,q)>\tau\}|/N$ with $\tau=0.8$, which the LLM may
accept or revise. \textbf{S10} fully decouples the pools: Part~1 retrieves the
semantic top-$K$ from the entire corpus for high-quality examples, and Part~2
draws a separate random sample for an unbiased prevalence estimate (again with
the threshold hint).

\paragraph{Query reformulation methods (S12, S14, S15).}
These keep S10's decoupled design but upgrade the Part~1 retrieval. \textbf{S12}
applies Rocchio pseudo-relevance feedback~\cite{rocchio1971relevance}: an initial
query retrieves the top~50 comments, the top~5 are treated as pseudo-relevant and
their embeddings averaged into a centroid $\bar{d}$, and the query vector is
updated as $q' = \alpha\,q_0 + \beta\,\bar{d}$ with $\alpha=0.7$, $\beta=0.3$
before a final retrieval pass. \textbf{S14} performs TF-IDF query expansion: the
top-$M$ terms by TF-IDF weight over an initial retrieval pass are appended to the
query string, which is then re-embedded; a 360-configuration grid search on
held-out topics (optimizing P@6) selected an initial pool of 20, $M=3$, and a
final pool of 40. \textbf{S15} is HyDE~\cite{gao2023precise}: the LLM first writes
a short hypothetical public comment it expects would be relevant to the topic,
and the embedding of that synthetic comment -- rather than the query
string -- retrieves the real comments.

\subsection{Topics}
The 20 evaluation topics are: Housing; Sustainability; Public Safety;
Affordability; Transit Corridors and Parking; Zoning and Land Use; Public Health;
Policing; Local Economy; Quality of the Built Environment; Public Spaces and
Recreation; Election Administration; Incarceration and Crime History;
Homelessness; Public Works; Anti-racism; Utility Service;
Senior/Infant/Child/Teenager Care; Public Service; and Functional Democracy.